\documentclass[12pt,preprint]{aastex}
\usepackage{color}

\newcommand{\BPH}{BPH04}

\newcommand{\sm}[1]{\mbox{{\scriptsize #1}}}
\newcommand{\simle} {\,{}^<_{\sim}\,}
\newcommand{\simge} {\,{}^>_{\sim}\,}

\newcommand{\be}{\begin{equation}}
\newcommand{\ee}{\end{equation}}
\newcommand{\bea}{\begin{eqnarray}}
\newcommand{\eea}{\end{eqnarray}}
\newcommand{\bdm}{\begin{displaymath}}
\newcommand{\edm}{\end{displaymath}}
\newcommand{\bef}{\begin{figure}}
\newcommand{\eef}{\end{figure}}
\newcommand{\befone}{
  \begin{figure*}
  \centering
  \begin{minipage}{\textwidth}
  }
\newcommand{\eefone}{\end{minipage}\end{figure*}}

\newcommand{\cm}{\mbox{cm}}
\newcommand{\m}{\mbox{m}}
\newcommand{\km}{\mbox{km}}
\newcommand{\AU}{\mbox{AU}}
\renewcommand{\sec}{\mbox{s}}
\newcommand{\g}{\mbox{g}}
\newcommand{\G}{\mbox{Gauss}}
\newcommand{\kG}{\mbox{kGauss}}
\newcommand{\Msol}{\mbox{$M_{\sun}$}}
\newcommand{\Rsol}{\mbox{$R_{\sun}$}}

\newcommand{\au}{\mbox{AU}}
\newcommand{\K}{\mbox{K}}
\newcommand{\yr}{\mbox{year}}
\newcommand{\ys}{\mbox{years}}

\newcommand{\vx}{{\bf x}}

\newcommand{\dd}{{\mbox{d}}}

\def\epstwover@scaling{0.4}

\def\plottwover#1#2{
  \centering
  \leavevmode
  \includegraphics[height={\epstwover@scaling\textheight}]{#1} \hfil \,
  \includegraphics[height={\epstwover@scaling\textheight}]{#2} \hfil \,
}

\def\epsover@scaling{0.85}

\def\plotover#1{
  \centering
  \leavevmode
  \includegraphics[height={\epsover@scaling\textheight}]{#1} \hfil \,
}

\def\epstwo@scaling{0.48}

\def\plotfour#1#2#3#4{
  \centering
  \leavevmode
  \includegraphics[width={\epstwo@scaling\linewidth}]{#1} \hfil
  \includegraphics[width={\epstwo@scaling\linewidth}]{#2} \hfil
  \includegraphics[width={\epstwo@scaling\linewidth}]{#3} \hfil
  \includegraphics[width={\epstwo@scaling\linewidth}]{#4} \hfil
}

\shorttitle{Outflows and jets from collapsing magnetized cloud cores} 
\shortauthors{R. Banerjee and R. E. Pudritz}

\begin{document}

\title{Outflows and Jets from Collapsing Magnetized Cloud Cores}

\author{Robi Banerjee$^1$ and Ralph E. Pudritz$^{1,2}$} 
\affil{$^1$ Department of Physics and Astronomy, McMaster University,
  Hamilton, Ontario L8S 4M1, Canada \\ 
  $^2$ Origins Institute, McMaster University, Arthur Bourns Bldg 241,
  Hamilton, Ontario L8S 4M1, Canada}

\begin{abstract}

Star formation is usually accompanied by outflow phenomena. There is
strong evidence that these outflows and jets are launched from the
protostellar disk by magneto-rotational processes. Here, we report on
our three dimensional, adaptive mesh, magneto-hydrodynamic simulations
of collapsing, rotating, magnetized Bonnor-Ebert-Spheres whose
properties are taken directly from observations. In contrast to the
pure hydro case where no outflows are seen, our present simulations
show an outflow from the protodisk surface at $\sim 130 \, \AU$ and a
jet at $\sim 0.07 \, \AU$ after a strong toroidal magnetic field build
up. The large scale outflow, which extends up to $\sim 600 \, \AU$ at
the end of our simulation, is driven by toroidal magnetic pressure
(spring), whereas the jet is powered by magneto-centrifugal force
(fling). At the final stage of our simulation these winds are still
confined within two respective shock fronts. Furthermore, we find that
the jet-wind and the disk-anchored magnetic field extracts a
considerable amount of angular momentum from the protostellar
disk. The initial spin of our cloud core was chosen high enough to
produce a binary system. We indeed find a close binary system
(separation $\sim 3 \, \Rsol$) which results from the fragmentation of
an earlier formed ring structure. The magnetic field strength in these
protostars reaches $\sim 3 \kG$ and becomes about $3 \, \G$ at $1 \,
\AU$ from the center in agreement with recent observational results.

\end{abstract}


\section{Introduction}
\label{sec:intro}

One of the earliest and most significant manifestations of star
formation is the appearance of jets and outflows in the early stages
of gravitational collapse \citep[for a review see][]{Bachiller96,
Andre00}. Theoretical work by~\citet{Blandford82}
and~\citet{Pudritz83} first established the idea that these outflows
originate from protostellar disks threaded by magnetic fields.  The
accretion flow in protostellar disks advect their magnetic field
inward while disk rotation leads to field configurations that flings
the gas off the disks.  Subsequent computational efforts
\citep{Shibata85, Uchida85b, Shibata86, Stone92} found large-scale,
magnetically launched outflows from collapsing disks around young
stellar objects.  Detailed magneto-hydrodynamical (MHD) simulations
showed that highly collimated, high velocity jets are driven by a
Keplerian accretion disk when threaded with magnetic fields
\citep[e.g.,][]{Ouyed97}. The large, comprehensive, theoretical and
computational literature that has developed in this field over the
last decade is discussed in recent reviews \citep[e.g.,][]{Konigl00,
Heyvaerts03, Pudritz03}. Current observations confirm that jets
rotate, and carry off angular momentum from their underlying disks
\citep[e.g.,][]{Bacciotti02}.

One of the most difficult aspects of star formation is to include both
magnetic fields and (the resulting) outflows into the problem. The
point is that protostellar disks are themselves the result of the
gravitational collapse of rotating cloud cores. Therefore, there
should be a profound link between collapse and outflow. Progress has
recently come from the advent of sophisticated MHD codes. One the
first simulations of collapsing magnetized cloud cores showed that a
magnetically driven outflow is launched after strong toroidal magnetic
field component is dynamically build up during the collapse of the
rotating molecular cloud~\citep{Tomisaka98}.

More recent simulations of collapsing magnetized cloud cores include
the work of~\citet{Tomisaka02, Boss02, Matsumoto04, Hosking04,
Machida04, Ziegler05, Machida05a, Machida05b}. Differences between
these approaches are based on numerical techniques and/or simulated
physics. For instance, a static spherical coordinate mesh is used
by~\citet{Boss02} where magnetic fields are treated in an approximate
way, whereas a Lagrangian magnetohydrodynamic approach is used
by~\citet{Hosking04} to address the question of fragmentation of a
magnetized protodisk. These groups arrive at different
conclusions. The results from the new AMR code NIRVANA
by~\cite{Ziegler05} confirmed the findings of~\citet{Hosking04} that
magnetic fields tend to stabilize the protodisk and prevent
fragmentation. Extensive nested grid simulations performed
by~\citet{Machida05a, Machida05b} focused also on fragmentation and
binary formation and concluded with magnetic flux-spin relations in
different regimes that controls fragmentation.

The issue of initial conditions for gravitational collapse and star
formation is important, but is, as yet not fully explored in 3D MHD
collapse simulations. So far, there is only one simulation of a
collapsing magnetized, cloud core that is embedded in a (low density)
environment \citep{Matsumoto04}. However, these authors do not focus
on the collapse--outflow connection but rather on the important
dynamical evolution and relation of the cloud angular momentum and the
field geometry. Other simulations are done for an assumed infinite
cylindrical cloud \citep[e.g.,][]{Tomisaka02, Nakamura03,
Machida04}. Such a cylindrical cloud does not go through an initial
phase of magnetic braking. Magnetic braking, even before the collapse
phase, has a significant effect on the initial rotation profile and
the angular momentum of the cloud \citep{Mouschovias80}.  Here, we
place the cloud core in a warm, low density, environment which is
entirely threaded with a background magnetic field. Such cloud cores
are frequently seen in observations~\citep[see e.g.][]{Ward02}. Moreover, the
presence of magnetic fields in molecular cloud cores is well justified
by observations \citep{Crutcher99}.

A far better understood aspect of star formation, at least in
principle, is the role of cooling. Most simulations are either based
on a pure isothermal equation of state (EOS) or an
isothermal-polytropic EOS where the EOS is locally switched whenever
the density exceeds a certain critical value. The latter approach,
which accounts for the different cooling mechanisms during the
collapse phase \citep[e.g. see][]{Larson03}, is not without its own
difficulties. First, changing the EOS from an isothermal system to a
polytropic system corresponds to a change from a system with an
infinite heat reservoir to a system with a finite reservoir and is
difficult to achieve numerically. Second, cooling by adiabatic
expansion of the gas is not correctly taken into account for a
polytropic index other than $5/3$~\footnote{Usually, a large scale
outflow is accompanied by an expanding ``magnetic bubble'' where
adiabatic cooling might affect its efficiency.}. Therefore, in order
to achieve a useful step towards a realistic picture of collapsing
cloud cores, we incorporate cooling by molecular line emission in our
simulations in this paper. This allows us to capture some of the real
complexity of the shock system that arises during the collapse
phase. We point the reader to Sec.~\ref{sec:properties} (where we show
the effective equation of state as a function of the core density) for
further discussion on the necessity of incorporating additional
physics into collapse simulations.

In this paper we present a comprehensive and self-consistent picture
of the generation of outflow phenomena driven by dynamically enhanced
magnetic fields that are build up within collapsing cloud cores. Our
study is based on the results of 3D adaptive mesh refinement (AMR) MHD
simulations that resolves the collapse of the magnetized, rotating
cloud core configurations over 7 decades in physical scale. Here, we
focus on the results of our low mass simulation ($M_{\sm{cloud}} = 2.1
\, \Msol$) whose initial conditions are chosen to closely match those
of the observed Bonner-Ebert molecular cloud core, Barnard
68~\citep{Alves01}. We also performed pure hydro simulations (without
magnetic fields) with a very similar setup to that reported here
\citep*[][(\BPH)]{Banerjee04}. We report on the differences and
similarities of magnetic/non-magnetic systems within this work.


\section{Simulation model and initial conditions}
\label{sec:model}

Our initial conditions (see also~\BPH~for the hydro setup) resemble
the properties of the well studied Bok globule, Barnard 68. Extinction
measurements show that the density of this cloud core closely follows
a Bonnor-Ebert-profile \citep{Alves01} (for a brief review of the
Bonnor-Ebert-Sphere see e.g. \BPH). We adopt the values from these
measurements for our initial setup: core density $\rho_0 = 9.81\times
10^{-19} \,\g \, \cm^{-3}$, mass of the cloud core $M = 2.1 \, \Msol$,
radius $R = 1.25\times 10^4 \, \AU$ (dimensionless Bonnor-Ebert-radius
$\xi = 6.9$) , and gas temperature $T = 16 \, \K$. The sphere is
bounded by a warm, low density (density contrast $\delta = 10$),
ambient medium so that the pressure at the edge of the sphere and that
of the ambient gas match. Furthermore, we assume a solid body rotation
of the cloud core with an angular velocity of $\Omega = 1.89\times
10^{-13} \, \mbox{rad} \, \sec^{-1}$ which corresponds to $\Omega \,
t_{\sm{ff}} = 0.4$, where $t_{\sm{ff}} = 2.12\times 10^{12} \, \sec$
is the initial free fall time. We chose this value because
hydrodynamic simulations show that cores that spin at rates $\Omega \,
t_{\sm{ff}} \simge 0.1$ fragment into rings and then binaries
\citep[e.g.,][\BPH]{Matsumoto03}.

The initial magnetic field is setup to be parallel to the rotation
axis ($z$-axis) which threads the entire simulation box. To account
for a magnetic pressure enhancement during core formation we assume a
constant thermal-to-magnetic-pressure, $\beta = p/(B^2/8\pi)$, in the
equatorial plane with a value of $\beta = 84$ which gives a minimal
and maximal field strength of $B_{\sm{min}} = 3.4 \, \mu\G$ and
$B_{\sm{max}} = 14 \, \mu\G$, respectively. Recent simulations of
collapsing, magnetized Bonnor-Ebert-Spheres \citep{Matsumoto04} showed
that the rotation axis will be aligned with the magnetic field during
the collapse of the cloud core as the perpendicular (to the magnetic
field) component of the angular momentum is extracted more quickly
than the parallel component. Therefore, an aligned rotator, as in our
case, might be a ``natural'' configuration of magnetized cloud
cores. In order to induce the collapse of the modified
Bonnor-Ebert-Sphere above, we enhance the density by 10\% to overcome
the additional rotational and magnetic component. We observe
fragmentation in the self-gravitating disk that forms as a consequence
of the collapse. We not add $m = 1$ or $m = 2$ perturbations however
-- the numerical perturbations suffice.


We follow the evolution with the grid based 3D MHD code FLASH
\citep{FLASH00} which is build on a block structured adaptive mesh
refinement technique \citep{PARAMESH99}. This technique enables us to
obey the Truelove criteria \citep{Truelove97} wherein the mesh must be
fine enough to resolve the local Jeans length by at least 4 grid
points in order to prevent spurious (i.e. numerically induced)
fragmentation. In our simulations, we resolve the Jeans length by at
least 8 grid points. At the end of our simulation we reach 27
refinement levels (the highest refinement level at the beginning of
the simulation is 7) which corresponds to a minimal grid spacing of
$\Delta x = 4.66\times 10^9 \, \cm = 0.067 \, \Rsol$. This should be
compared to the length of our simulation box which is $2.5 \times
10^{18} \, \cm$. In this high resolution run we are able to study
details of the protostellar environment and the fragmentation to a
very close binary system (separation $\sim 3 \, \Rsol$). We performed
several lower resolution runs with different magnetic field strengths
and different angular velocities. The high resolution simulation
presented in this paper, ran for 44 hours on 32 processors on an Alpha
SC Server.

\subsection{Cooling}
\label{ssec:cooling}

As in our hydro simulations~(\BPH), we augmented the FLASH code with
the ability to cool the gas as it contracts using the self-consistent
radiative cooling calculations by~\citet{Neufeld93, Neufeld95}.  So
far, our simulations do not include the possible cooling by dust
grains coupled to the gas. Cooling of molecular gas by dust grains is
very efficient even in the regime $n > 10^{7.5} \, \cm^{-3}$ if the
gas is tightly coupled to grains \citep[see e.g.,][]{Goldsmith78,
Goldsmith01}. The critical density at which dust
cooling becomes inefficient is in the optical thick regime at a
density of about $10^{10} \, \cm^{-3}$ after which the core becomes
essentially adiabatic~\citep[see e.g.][]{Larson03} rather than the
value of $10^{7.5} \, \cm^{-3}$ which is set by the molecular cooling
scale. In~\BPH\, we found that the appearance of shocks results when
$t_{\sm{cool}} \sim t_{\sm{ff}}$. A prolonged isothermal collapse
phase (due to efficient dust cooling) would shift the appearance of
the first shock (see Sec.~\ref{sec:collapse}) to higher densities and
therefore smaller scales but would not change the basic physical
outcome. 

We derived the relation between the
critical density and the scale height (\BPH, Eq.~(21)) at which the shock
first appears as 
\be
r_{\sm{shock}} \sim 450 \, \au \, 
  \left(\frac{n_{\sm{crit}}}{10^{7.5} \, \cm^{-3}} \right)^{-1/2}
  \,. 
\ee
Therefore, we would expect the shock scale
to set in at about $25 \, \AU$ above the protodisk if the critical
density is $\sim 10^{10} \, \cm^{-3}$, compared to our findings which
show the shock appearance at $\sim 450 \, \AU$. Recent 1D
simulations by~\citet{Lesaffre05} including gas and dust cooling
indeed find a accretion shock at $\sim 10 \, \au$ which is close to
the value to which we extrapolate our own results if cooling by dust
is included. Simulations by others find also similar shock and/or
outflow structures \citep[e.g.,][]{Yorke95, Tomisaka02,
Matsumoto04}. Thus, even without explicit dust cooling in our code, we
can confirm that our results should scale. We will explicitly add dust
cooling in our future work.

\subsection{Ambipolar diffusion}
\label{ssec:ambiploar}

Another physical process which we have not yet accounted for is
ambipolar diffusion, the slipping of magnetic field lines relative to
neutrals in the weak ion-neutral coupling regime. While many papers
have been written about field limitation by ambipolar diffusion
\citep[e.g.,][]{Ciolek98, Li98, Desch01}, they have so far focused on
spherical or asymmetric configurations with purely poloidal magnetic
fields. Moreover, the calculations are done for highly idealized
conditions for the ionization balance in such systems. Consider first
the effect of rotation. A collapsing rotating core and disk will have
a maximum value of the toroidal field near the inner edge of the disk
(since the toroidal field on the rotation axis must necessarily
vanish) and will have a quadrupolar spatial geometry
(cf. Sec.~\ref{sec:disk_evolution}). Such toroidal field will exert a
pressure gradient force towards the disk mid-plane (in addition to the
pressure gradients that drives outflows away from the disk) as well as
towards the protostar(s). The latter effect may serve to stabilize the
protostellar field and significantly reduce the ambipolar diffusion
rate.
 
Recent numerical simulations have also shown that it is possible to
have a strong fossil field within a star that is stable and lies in
the range of $\simge 10^3 \, \G$ \citep{Braithwaite04}. Such a field
is stabilized by a ring of toroidal fields that is buried under the
stellar surface. This situation somewhat resembles our own simulation
results. Thus, the inclusion of full 3D field dynamics may yet show
that ambipolar diffusion does not necessarily exclude the formation of
a strong fossil field.

A second general point is that the disks that are formed during
collapse will generally have well ionized surface layers -- even if
cosmic rays are the only source for ionization. While ambipolar
diffusion should occur in the mid plane, the surface layers of the
disk may retain strong fields.

The first direct measurement by~\citet{Donati05} of magnetic fields in
the FU Ori accretion disk using Zeeman signatures indicate strong
magnetic fields very close to the protostar ($\sim 1\,\kG$ at $0.05 \,
\au$). As these authors make clear, the observed magnetic fields are
disk-fields and are not generated in the protostar. At least in this
case, it appears as if ambipolar diffusion is not effective enough to
prevent the trapping of strong magnetic fields in the surface layers
of the disk.

\section{Magnetic braking in the pre-collapse phase}
\label{sec:precollapse}

\bef
\plottwo{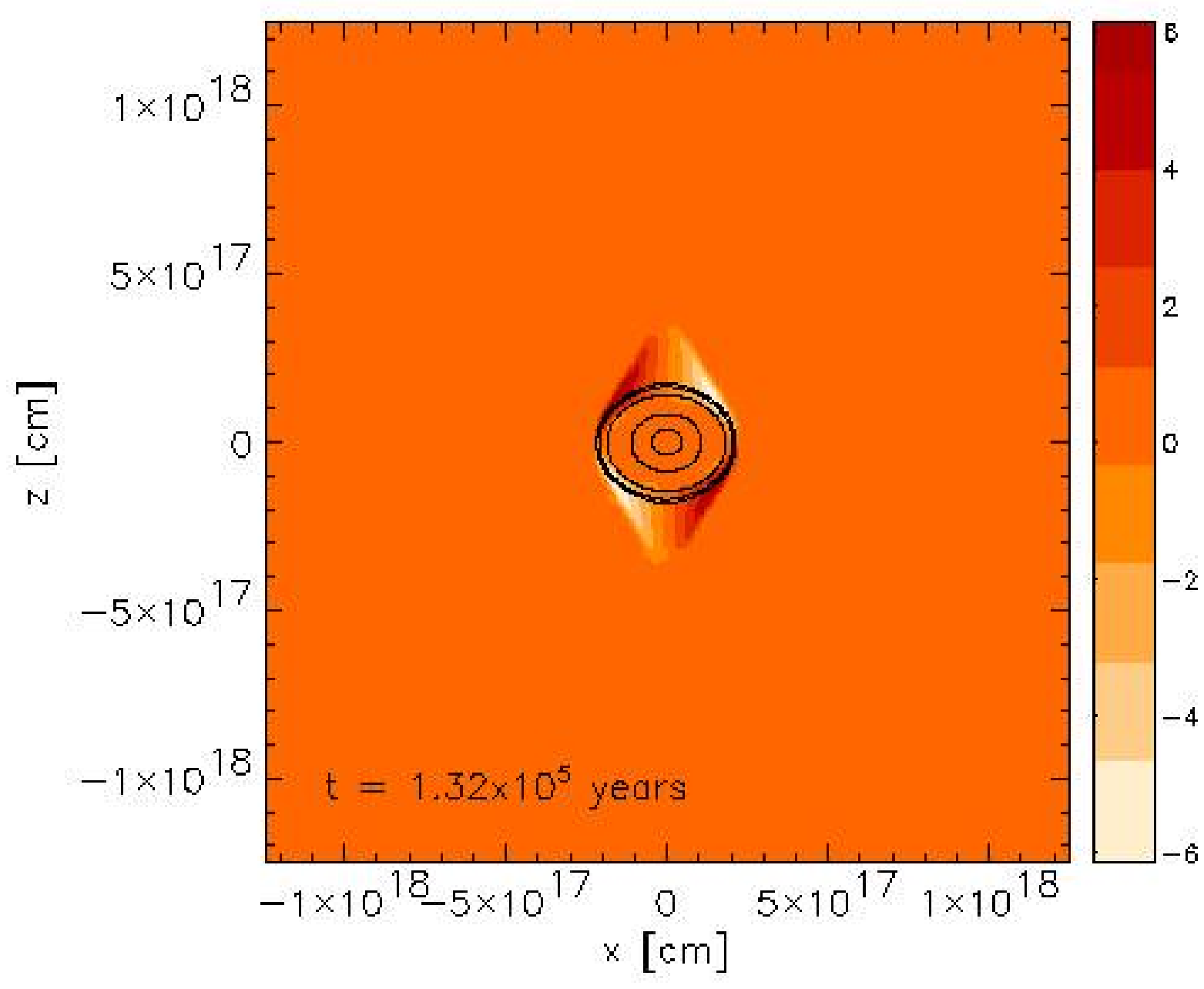}
        {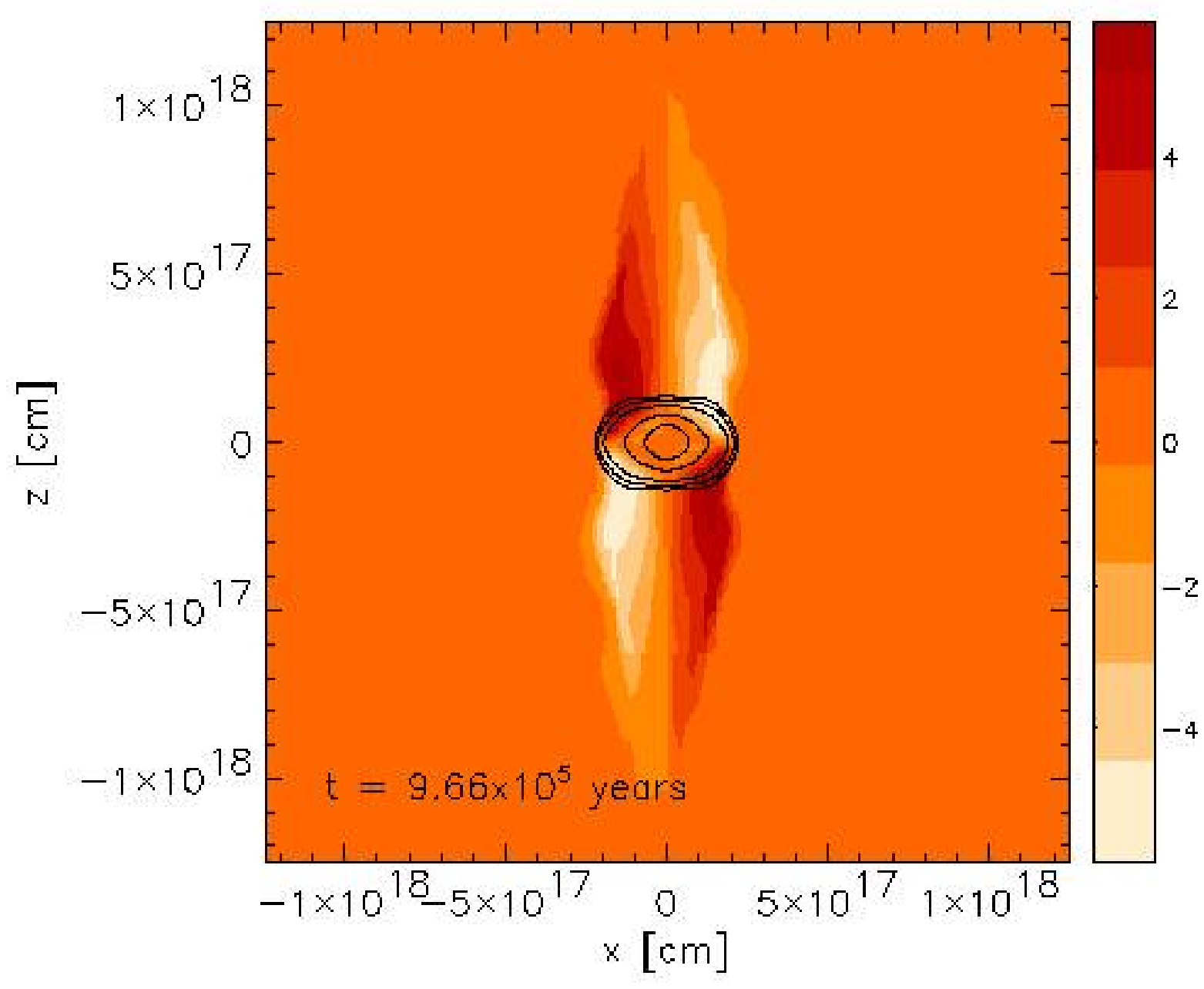}
\caption{Shows the torsional Alfv\'en waves that are launched into the
  non-rotating ambient medium (color scale in $\mu\G$). These waves
  move at the speed of the local Alfv\'en velocity ($v_{\sm{A}} = 486
  \, \m \, \sec^{-1}$) and transfer angular momentum from the
  rotating cloud core to the low density environment leading to a slow
  down of the cloud core \citep{Mouschovias80}. The snapshots which
  show the toroidal magnetic field (gray scale in $\mu\G$) and density
  (contour lines) are taken at $t = 1.3 \times 10^{5} \, \ys$ (left
  panel) and $t \sim 1\times 10^6 \, \ys$ (right panel),
  respectively. The density contour lines show the location of the
  Bonnor-Ebert-sphere in the simulation box, where the side length of
  simulation box is $2.5 \times 10^{18} \, \cm = 1.67 \times 10^5 \,
  \AU$).}
\label{fig:tor_Alfven}
\eef

A spinning, magnetized core undergoes significant magnetic braking
even before the collapse begins. In this early phase, some of the
core's initial angular momentum will be carried off by a flux of
torsional (Alfv\'enic) waves which transfer angular momentum from the
spinning sphere to the surrounding gas gradually spinning down the
sphere. The initial magnetic braking of a spinning cloud core was
studied analytically by~\citet{Mouschovias80} for idealized disk-like
rotors embedded in a non-rotating environment, and its appearance
provides an excellent test of the accuracy of our numerical methods as
well. That cores do not co-rotate with their environments has been
explicitly shown in many turbulent fragmentation simulations of core
formation. The system of oblique shocks that produces a whole
spectrum of core masses is also the likely origin of the spin
distribution of cores~\citep[see e.g.,][]{Tilley04}.

We present the results of our simulation of this pre-collapsing stage
in Fig.~\ref{fig:tor_Alfven} where the propagation of the torsional
Alfv\'en wave in the ambient medium is shown.  The spin down time of a
homogeneous disk embedded in a non-rotating environment with a
density contrast $\delta = \rho_{\sm{cloud}} / \rho_{\sm{ext}}$
is~\citep{Mouschovias80}
\be
\tau_{\sm{damp}} \approx \frac{Z \, \delta}{v_{\sm{A,ext}}} \, ,
\label{eq:damping_time}
\ee 
where $v_{\sm{A,ext}} = B/\sqrt{4\pi\,\rho_{\sm{ext}}}$ is the
Alfv\'en velocity in external medium, and $Z$ is the disk
height. For our setup, this gives a spin down
time of $\sim 10^6 \, \ys$, which is in good agreement with our
numerical results shown in Fig.~\ref{fig:tor_Alfven}.

During the pre-collapse phase, the core's rotation is slowed by a
factor of $2$ by this magnetic braking mechanism which gives $\Omega
\, t_{\sm{ff}} \approx 0.2$ at the time when the major collapse gets
underway. The amount of rotation prior to the runaway collapse has an
important effect upon the formation of the disk since it determines
its size, fiducial density and hence its ability to cool and fragment
\citep[e.g.,][\BPH]{Matsumoto03}.  In particular, for $\Omega \,
t_{\sm{ff}} \simge 0.2$ a purely hydrodynamic disk fragments into a
ring which further fragments into a binary. We also note that in order
to generate sufficiently strong toroidal magnetic fields to get
magnetically driven outflows, the spin down time must not be too
small. Our simulations with stronger magnetic fields and/or slower
initial rotation showed no, or only a weak outflow as the build-up
toroidal magnetic field component is not strong enough to power a
wind. By comparing the damping time of Eq.~(\ref{eq:damping_time}) to the
initial orbital time of the sphere one can estimate whether magnetic
braking spins down the sphere too quickly. The condition that the
spin-down time not be too short gives~\citep[cf. Eq.~(18a)
in][]{Mouschovias80}:
\be 
\tau_{\sm{damp}} > 2\pi \, \Omega^{-1} \, ,
\label{eq:damp_rot}
\ee
Using $Z \approx c_{\sm{s}}/\sqrt{4\pi\,G\,\rho_0}$ this relation can
be rewritten as 
\be
\frac{\left(\delta \, \beta\right)^{1/2}}{\sqrt{3}\,\pi} \,
  t_{\sm{ff}} \, \Omega > 1
\label{eq:rot_beta}
\ee
(in our case $t_{\sm{ff}} \, \Omega \, \sqrt{\delta\,\beta/3}/\pi =
2.13$). Simulations where condition Eq.~(\ref{eq:rot_beta}) is not or
only barely fulfilled result in a more spherical collapse with less
pronounced disks and had no tendency to fragment.  We conclude from
these results that modest to strong magnetic fields stabilize the
cloud core and prevent fragmentation during the collapse (see also
the discussion in Sec.~\ref{sec:disk_evolution}). We note separately
that a measure of the toroidal field under these circumstances is
given by the expression~\citep[][Eq.~(17a)]{Mouschovias80}
\be
\frac{B_{\phi}}{B_z} = \frac{R \, \Omega}{{v_{\sm{A,ext}}}} \, ,
\label{eq:pitch}
\ee
where $B_z$ is the initial, homogeneous, magnetic field. For our
initial setup $B_{\phi}/B_z$ (Eq.~(\ref{eq:pitch})) becomes larger than
one at a disk radius of $1.14\times 10^{17} \, \cm$.

\section{The initial collapse phase}
\label{sec:collapse}

Initially, the sphere is close to a hydrostatic equilibrium so that
the runaway collapse starts only after a substantial amount of the
additional rotational support is extracted by magnetic braking
\citep[see also,][]{Basu94, Basu95a, Basu95b}. Here, the time from the
beginning of the simulation to the start of the runaway collapse is
$1.8\times 10^6 \, \ys$\footnote{Henceforth, times given in this paper
are measured form the beginning of the runaway collapse, i.e. $t =
t_{\sm{sim}}-t_0$, with $t_0=5.665\times 10^{13} \, \sec = 1.8 \times
10^6 \, \ys$}.

The first phase of the collapse of magnetized Bonnor-Ebert-Spheres
proceeds very similarly to the non-magnetized cases (see also \BPH):
initially the rotating cloud core collapses from outside-in
isothermally so long as molecular cooling is efficient enough to keep
the core on its initial temperature. At densities of about $n \sim
10^{7.5} \, \cm^{-3}$, cooling becomes inefficient and the dense core
starts to heat. At this point, a thick disk begins to form. The
infalling material undergoes a shock at a height of $\sim 450 \, \AU$
above and below the disk plane (if dust is included, we expect this
scale to be at about $25 \, \au$). In this early stage of the collapse
($50 - 60 \times 10^3 \, \ys$ into the collapse), these shock fronts
move steadily towards the center of the core. Not until the core
density exceeds $\sim 10^{10} \, \cm^{-3}$ is material falling onto
the core much affected by the magnetic field as the magnetic pressure
is still smaller than the thermal pressure ($\beta >
1$). Note that a general prediction of the density at
which the outflow will be launched will involve the knowledge of the
magnetic field structure and strength as well as the rotation velocity
of the first core. The non-linear interplay between the magnetic field
and the self-gravitational collapse of a rotating cloud hampers this
task. Nevertheless, we find that the outflow starts well into the
collapse and after the first core has formed.

\section{Onset of large scale outflow}
\label{sec:outflow}

\bef 
\plottwover{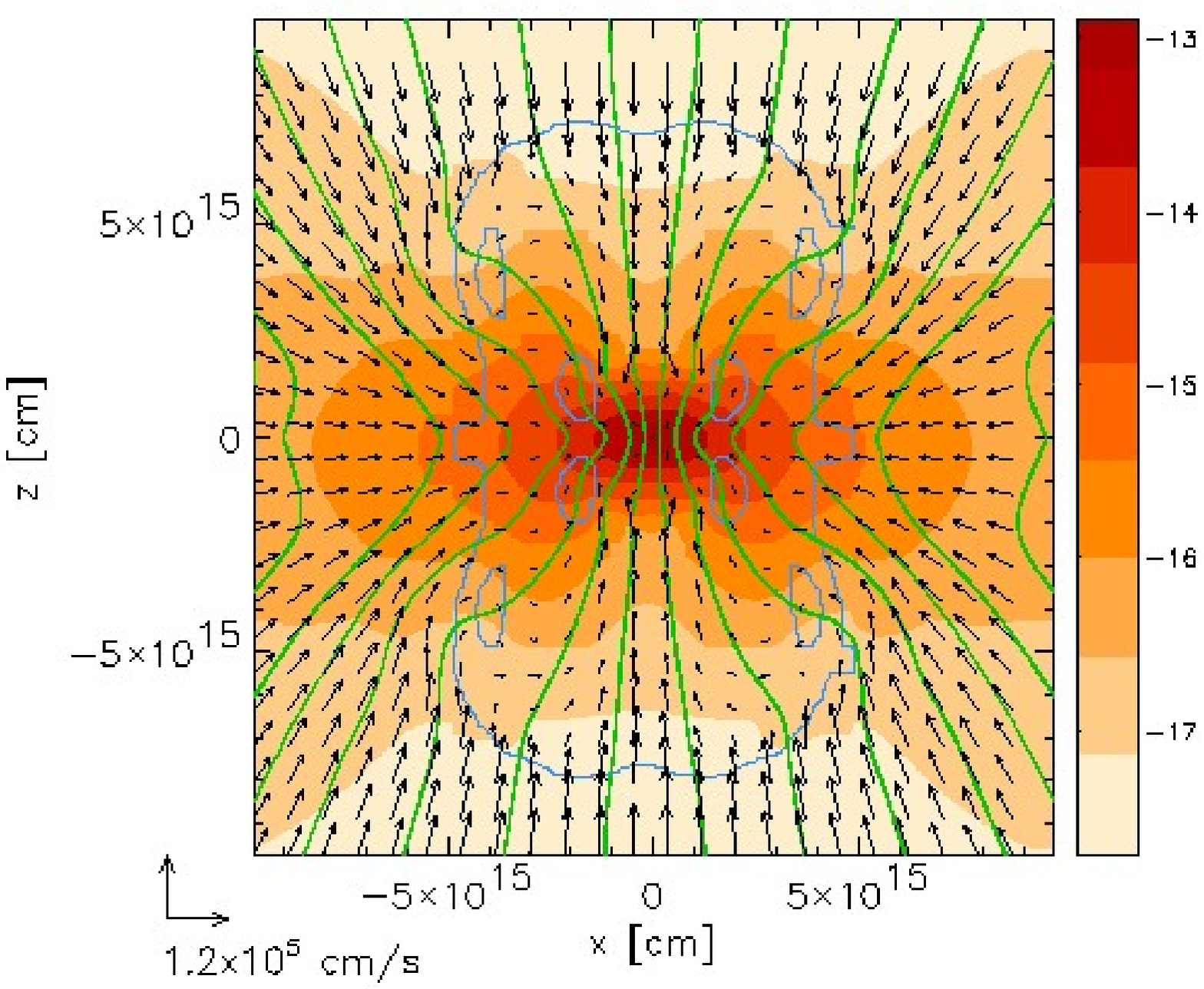}{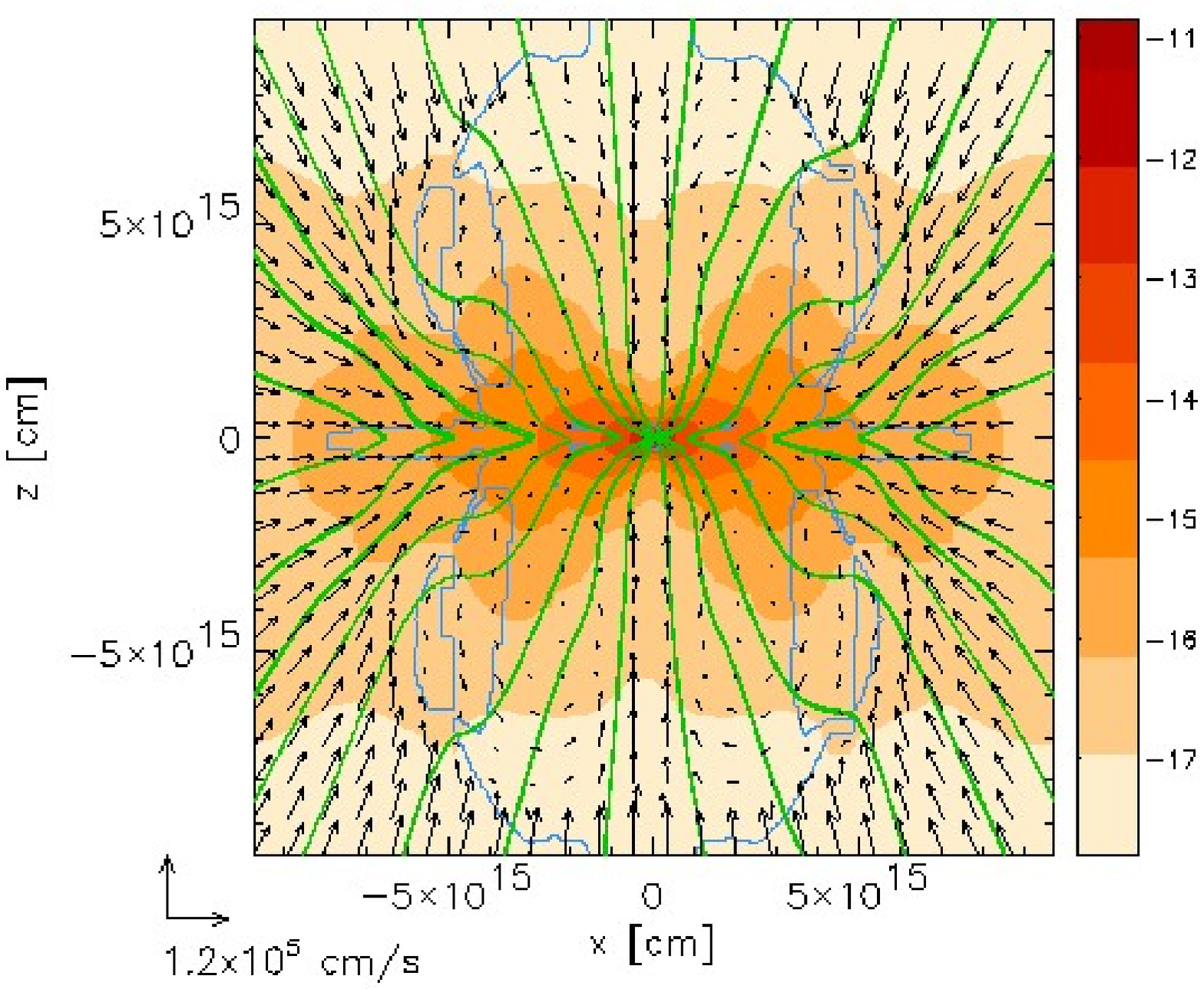}
\caption{Onset of the large scale outflow.The snapshots show the time
  evolution of the magnetically driven winds which are launched during
  the star formation process. The upper panel shows the collapsing
  stage at $t = 6.81\times 10^{4} \, \ys$ shortly after the onset of
  the outflow and the lower panel shows the situation $1430 \, \ys$
  later, at the end of our simulation, when the outflow is clearly
  visible. The magnetic pressure drives a ``bubble'' which is
  surrounded by shock fronts and reverses the gas flow. Here, the
  typical outflow velocity is $v_z \sim 0.4 \, \km \, \sec^{-1}$ at $z
  = 400 \, \au$. The color scale shows the gas density distribution
  (logarithmic scale in $\g \, \cm^{-3}$), the vector field reflects
  the velocity flow, the blue contour lines mark the Alfv\'en
  surfaces, and the magnetic flux surface (($B_z$, $B_x$) components)
  is drawn in green.}
\label{fig:outflow}
\eef

A strong toroidal magnetic field component builds up by winding
magnetic field lines as the core's angular velocity increases during
its contraction phase. By time the central density reaches $\sim
10^{10} \, \cm^{-3}$, the magnetic pressure from the toroidal field
component has become strong enough to prevent the shock fronts below
and above the disk plane from moving towards the center. Now, material
inside the magnetized bubble is pushed outward leading to a large
scale outflow. The onset of this large scale magnetic tower outflow
and the situation at the end of our simulation are shown in the two
snapshots in Fig.~\ref{fig:outflow} wherein we present a 2 pixel wide,
2D slice through the center of the simulation box. Such large scale,
low velocity outflows are also seen in other collapse simulations
\citep[e.g.,][]{Tomisaka98, Tomisaka02, Allen03b}.

This collimated bipolar outflow can be understood in terms of a
magnetic tower \citep{LyndenBell03, Kato04} that consists of an
annulus of highly wound magnetic field lines that pushes into the
ambient pressure environment. The toroidal magnetic field component
that is continuously produced by the rotating disk acts like a
compressed spring which lifts some material off the disk surface and
sweeps up material in the external medium.  Substantial pressure is
needed to trap the toroidal field, allowing it to wind up and push
into this region. Solutions show~\citep{LyndenBell03} that the
expansion of a tower grows linearly with time in the case of an
uniform external pressure and accelerates in the case of a decreasing
pressure profile. The acceleration of a tower may be slowed or even
reversed in the presence of ram pressure. Taken together, it appears
natural that a tower flow be driven from a disk within the
high-pressure region inside the first shock.

At this stage where the large scale outflow begins to sweep up
material the magnetic pressure becomes stronger than the thermal
pressure ($\beta \sim 0.1 - 1$) and the interior of the ``magnetic
bubble`` cools further by adiabatic expansion. Such an outflow may be
the origin of the molecular flow that is seen from all young stellar
objects (YSOs) \citep[e.g.,][]{Uchida85b}. Measurements of CO emission
lines of outflows from young stellar objects indicate higher
velocities, but our simulations show the star formation phase in a
very early stage wherein the central mass of the protostar is still
tiny. Since the outflow velocity is related to the escape speed
\citep[see][for a review]{Pudritz03}, this is the expected result.
The outflow speed will increase with time as central stellar mass
grows.  In Fig.~\ref{fig:outflow} one can see that the outflow
velocity exceeds the poloidal Alfv\'en velocity where the outflow is
the fastest. The outflow forces the region enclosed by the outer shock
fronts to expand and the shock fronts are moving outward. By the end
of our simulation the shock fronts are pushed to a disk height of
$\sim 600 \, \au$ and would presumably continue to rise. As already
mentioned, this may be the origin of bipolar outflows that are
associated with all young stellar objects.

We point out that the collimation of the large scale outflow is not
due to the initially uniform field which extends to infinity but
rather to the dynamically built up field structure which provides hoop
stresses to confine the outflow. The fact that toroidal field
component dominates the poloidal component in the outflow region shows
that the large scale outflow is driven by (toroidal) magnetic pressure
and confined by by the same toroidal field structure as shown
in~\citet{LyndenBell03}.

\section{The onset of the disk jet}
\label{sec:jet}

\bef
\plottwover{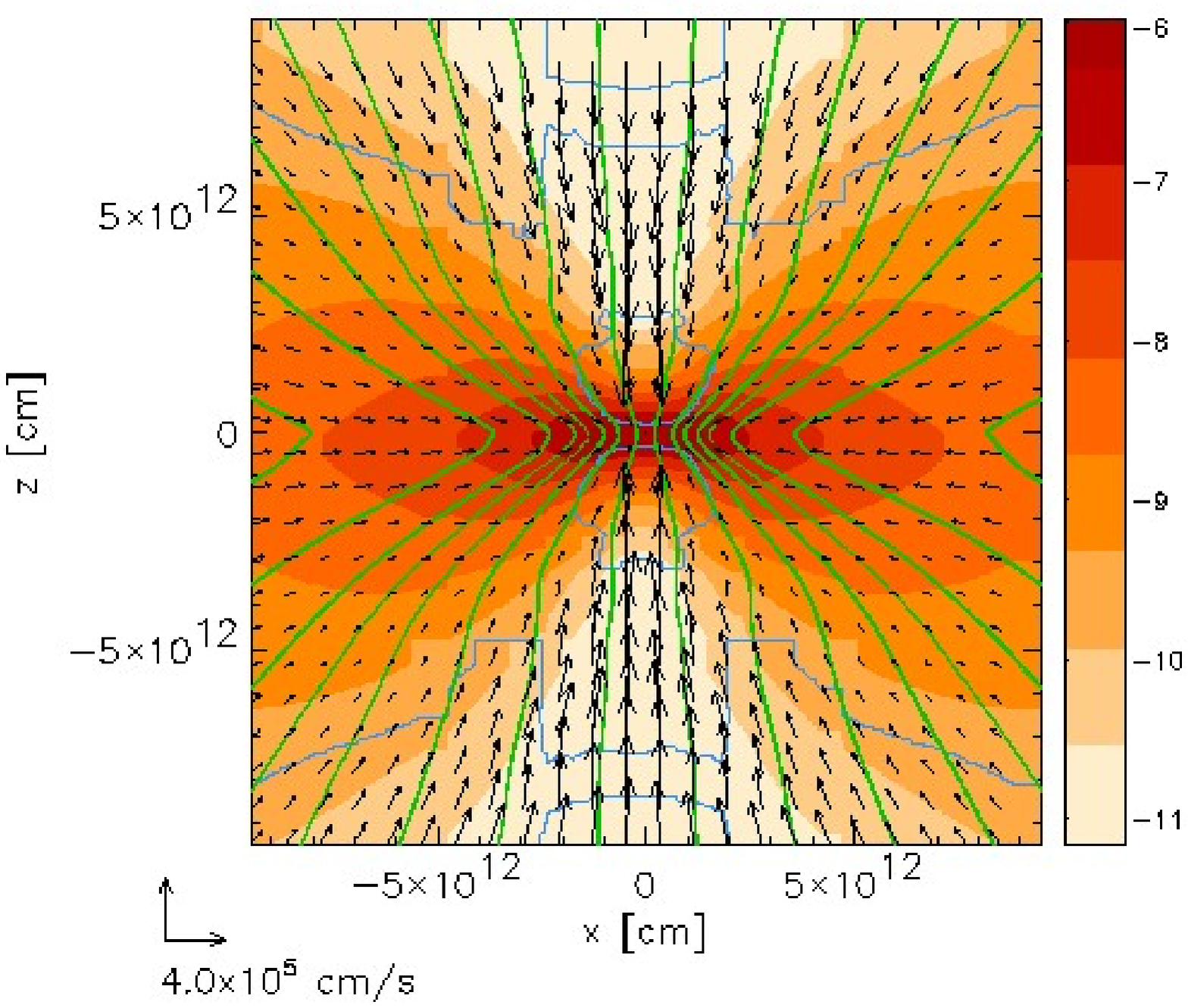}
	   {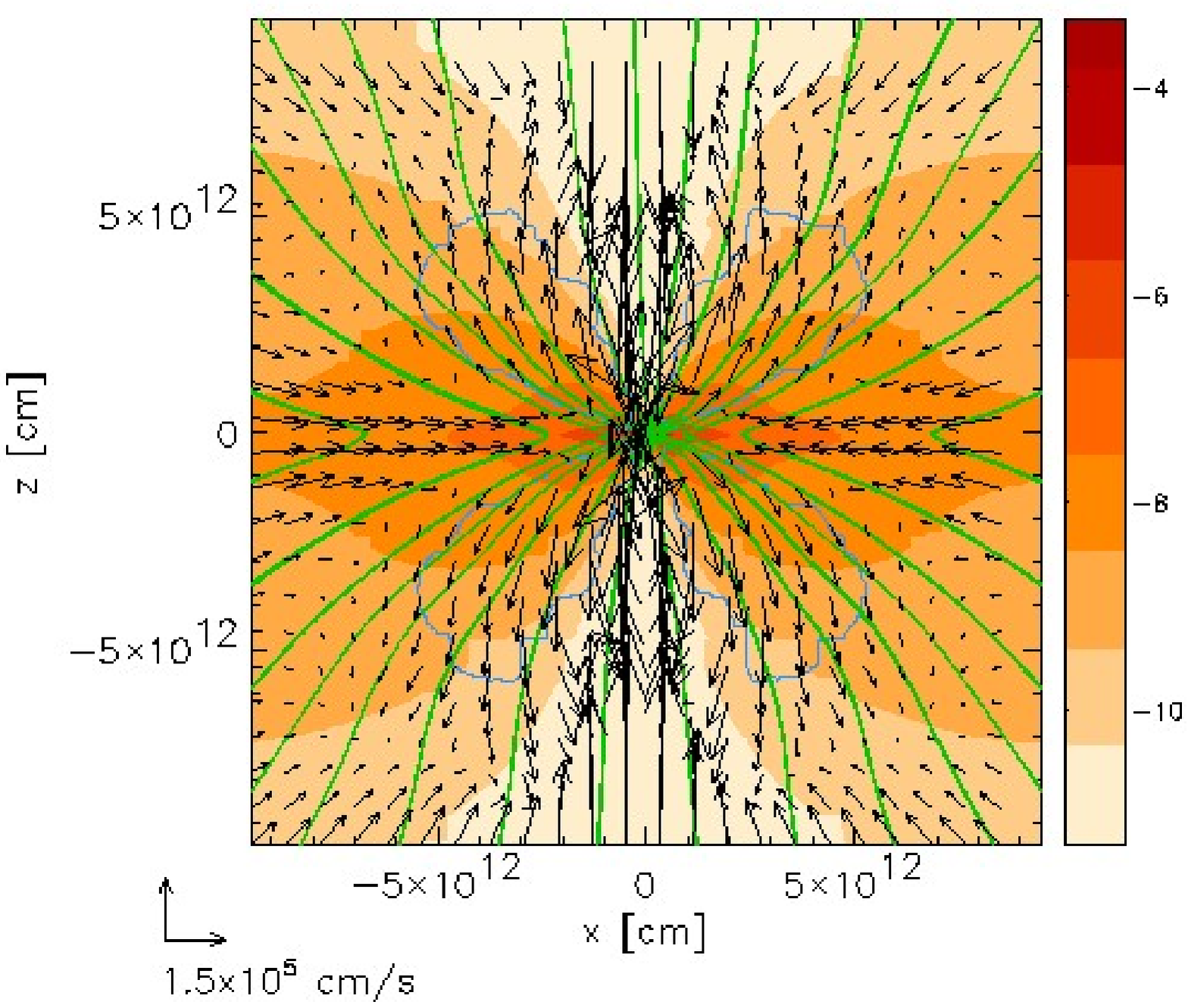}
\caption{Onset of the jet. The snapshots show the inner structure of
  the protostar and pre-stellar disk up to $\sim 0.7 \, \au$ (zoom
  factor 1000 compared to Fig.~\ref{fig:outflow}). Only 5 months
  ($1.258\times 10^7 \, \sec$) after the upper snapshot is taken (at
  $t = 6.89\times 10^{4} \, \ys$) jets above and below the pre-stellar
  disk are centrifugally driven by the strong magnetic field.  The jet
  velocities reach $\sim 3 \, \km \, \sec^{-1}$ at $0.4 \, \au$. The
  color coding and contour lines are described in
  Fig.~\ref{fig:outflow}.}
\label{fig:jet}
\eef

An even more dramatic outflow phenomenon erupts from the interior
regions of the disk, in the deepest part of the gravitational
potential well generated by the assembling protostar.  In
Fig.~\ref{fig:jet}, we show two snapshots of the disk and surrounding
infalling region focused down to a scale of $\sim 0.7\,\au$.  The
upper panel shows the collapse of material that is still raining down
onto the disk at time of $t = 6.8\times 10^4 \, \ys$.  In comparison
with the outer regions of the disk, the magnetic field lines towards
the disk interior have been significantly distorted as they are
dragged inwards by the disk's accretion flow.  They take the
appearance of a highly pinched-in, hour-glass. This configuration is
known to be highly conducive to the launch of disk winds
\citep{Blandford82, Pudritz83, Lubow94, Ferreira97}: magnetic field
lines threading the disk with an angle with the vertical that is
greater than $30^\circ$ are able to launch a centrifugally driven
outflow of gas from the disk surface. Our simulations clearly confirm
this picture as the angles of the magnetic field lines with the
vertical axis that are much greater than $30^\circ$.

Five months later in our simulation, a jet can clearly be seen to
leave the disk surface inside a spatial scale of $3 \times 10^{12} \,
\cm$.  Moreover, this disk wind achieves super-Alfv\'enic velocities
above which it begins to collimate towards the outflow axis. This jet
is much more energetic than the magnetic tower outflow at $1000$ times
larger scales (cf. Figure~\ref{fig:outflow}).

Similar to the large scale outflow, this jet is confined between shock
fronts which result from the non-isothermal EOS due to inefficient
cooling (for the evolution of the effective EOS see
Sec.~\ref{sec:summary}). The appearance of these shocks seems to
encourage flow reversal, possibly because the high pressure inside the
post-shock region.

\bef
\plotover{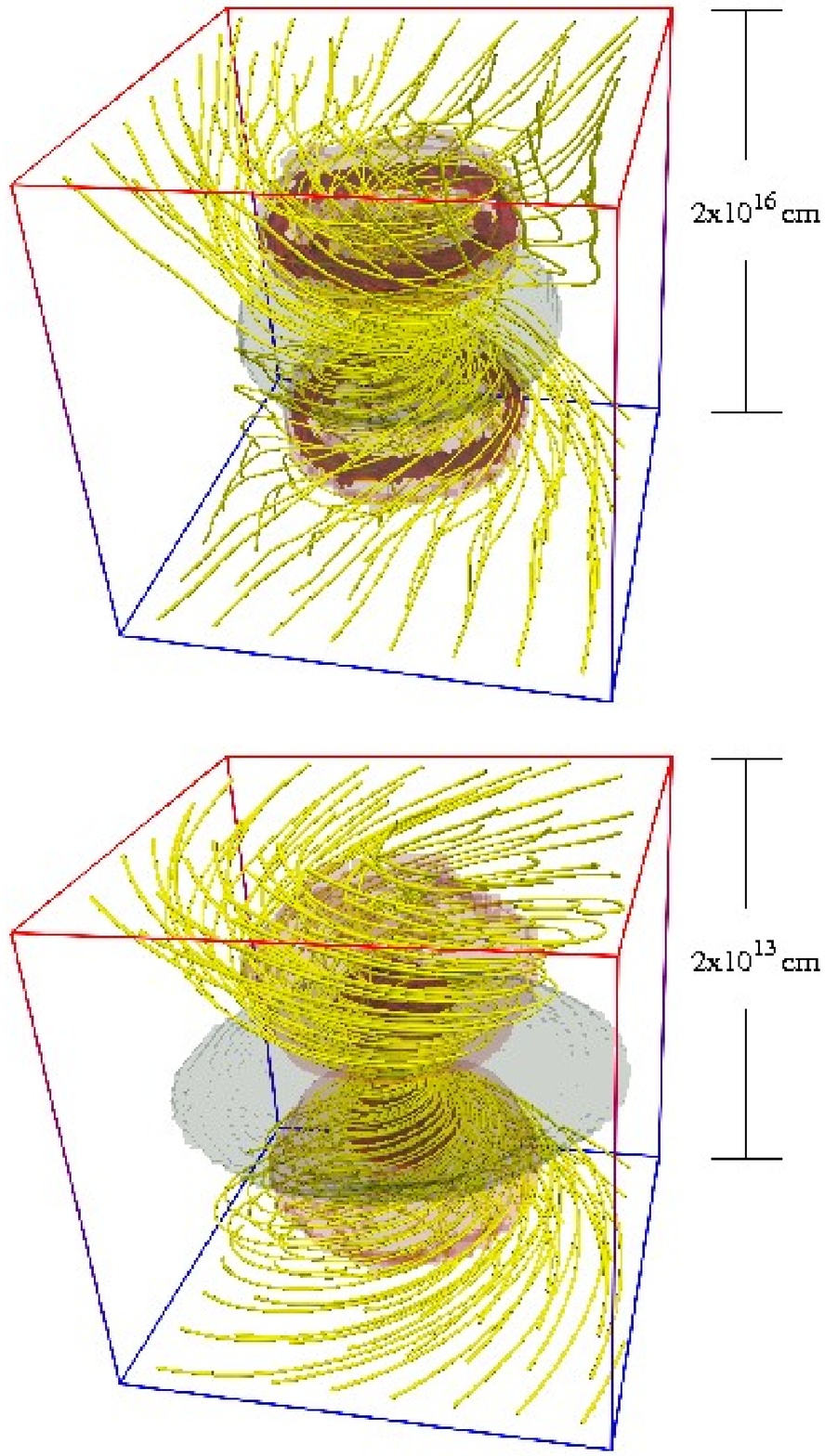}
\caption{Magnetic field line structure, outflow and disk. The two 3D
  images show the magnetic field lines, isosurfaces of the outflow
  velocities and isosurfaces of the disk structure at the end of our
  simulation ($t \simeq 7\times 10^{4} \, \ys$) at two different
  scales: the upper images refers to the scale shown in
  Fig.~\ref{fig:outflow}, (large scale outflow, side length of the box
  $L = 1.95\times 10^{16} \, \cm$) and the lower images shows the jet
  launching region of Fig.~\ref{fig:jet} ($L = 1.91\times 10^{13} \,
  \cm$). The isosurfaces of the upper panel refer to velocities
  $0.18\,\km\,\sec^{-1}$ (light red) and $0.34\,\km\,\sec^{-1}$ (red)
  and a density of $2\times 10^{-16} \, \g \, \cm^{-3}$ (gray) whereas
  the lower panel shows the isosurfaces with velocities
  $0.6\,\km\,\sec^{-1}$ (light red) and $2\,\km\,\sec^{-1}$ (red) and
  the density at $5.4\times 10^{-9} \, \g \, \cm^{-3}$ (gray).}
\label{fig:mag_flines}
\eef

In Figure~\ref{fig:mag_flines}, we show two snapshots of the three
dimensional structure of the magnetic field lines each taken at the
end of our simulation. The upper panel shows the 3D field structure of
magnetic field lines on the scale of the large scale outflow in
Figure~\ref{fig:outflow}, while the lower panel shows the same for the
jet in Figure~\ref{fig:jet}. The magnetic field lines in both
snapshots are always swept backwards in a rotating outflow - an affect
that arises from the mass loading of these flows. Nevertheless, there
is a difference in the field line structure - those in the upper panel
are almost parallel to the outflow axis outside the large scale
outflow region, while the field lines in the jet region have a strong
component parallel to the protodisk. The latter field structure
indicates the jet is magneto-centrifugally powered wherein the field
lines act as ``lever arms'' flinging material off the disk surface,
whereas the large scale outflow is powered by magnetic pressure in
which the magnetic field lines behave like the release of a compressed
spring.

\section{Ring fragmentation in the disk -- the formation of a
  proto-binary system} 
\label{sec:disk_evolution}

\bef \plotfour{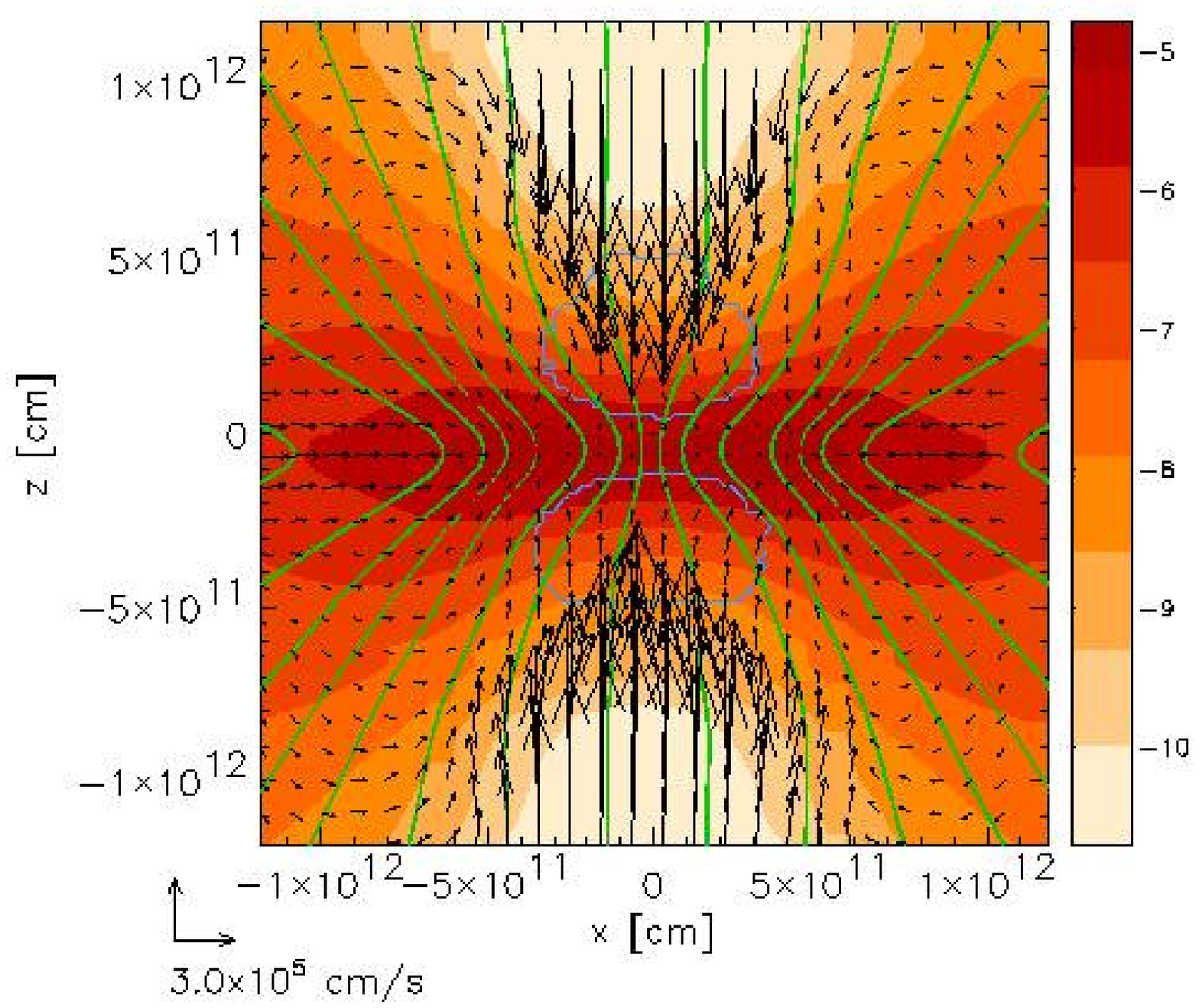} {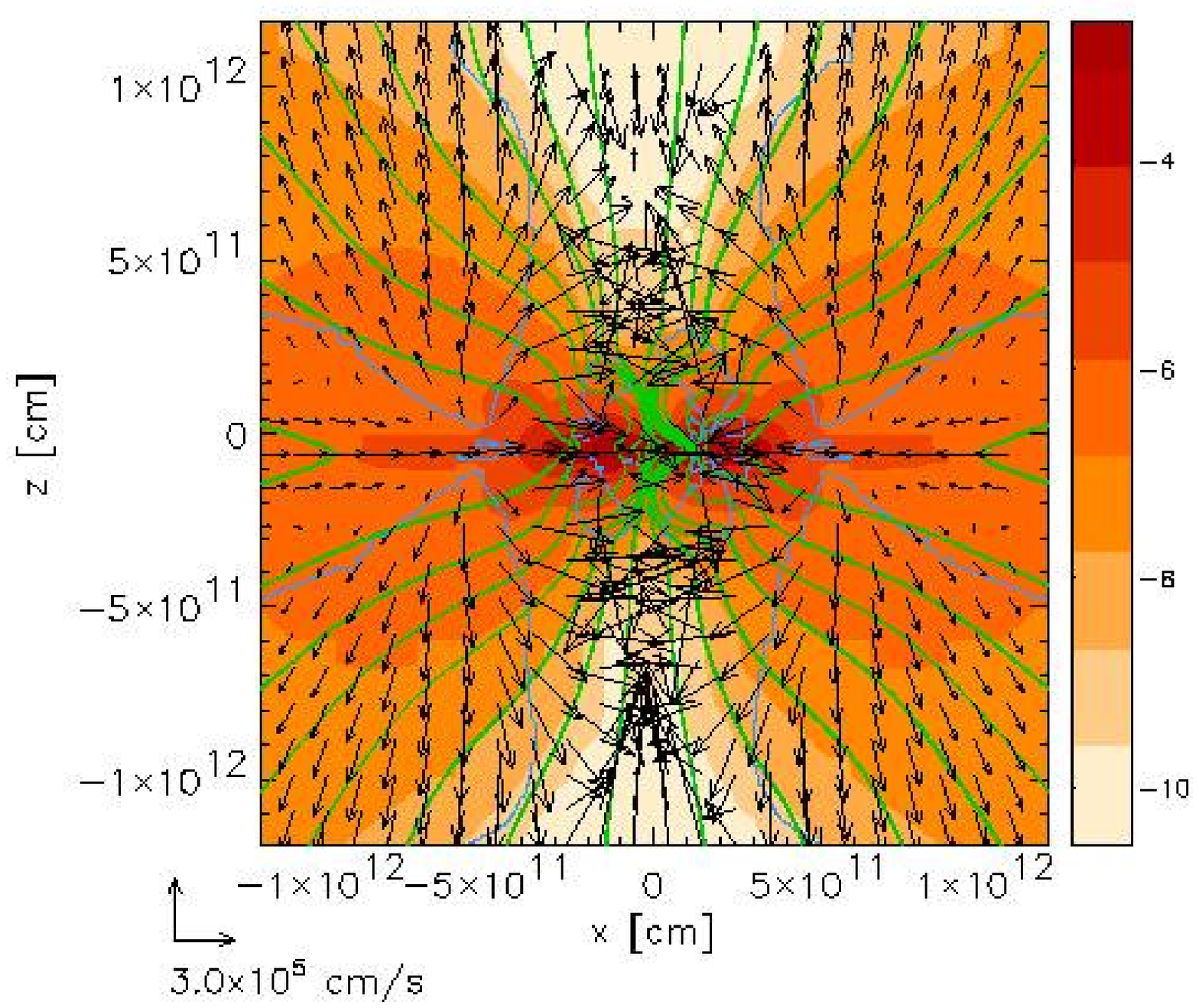} {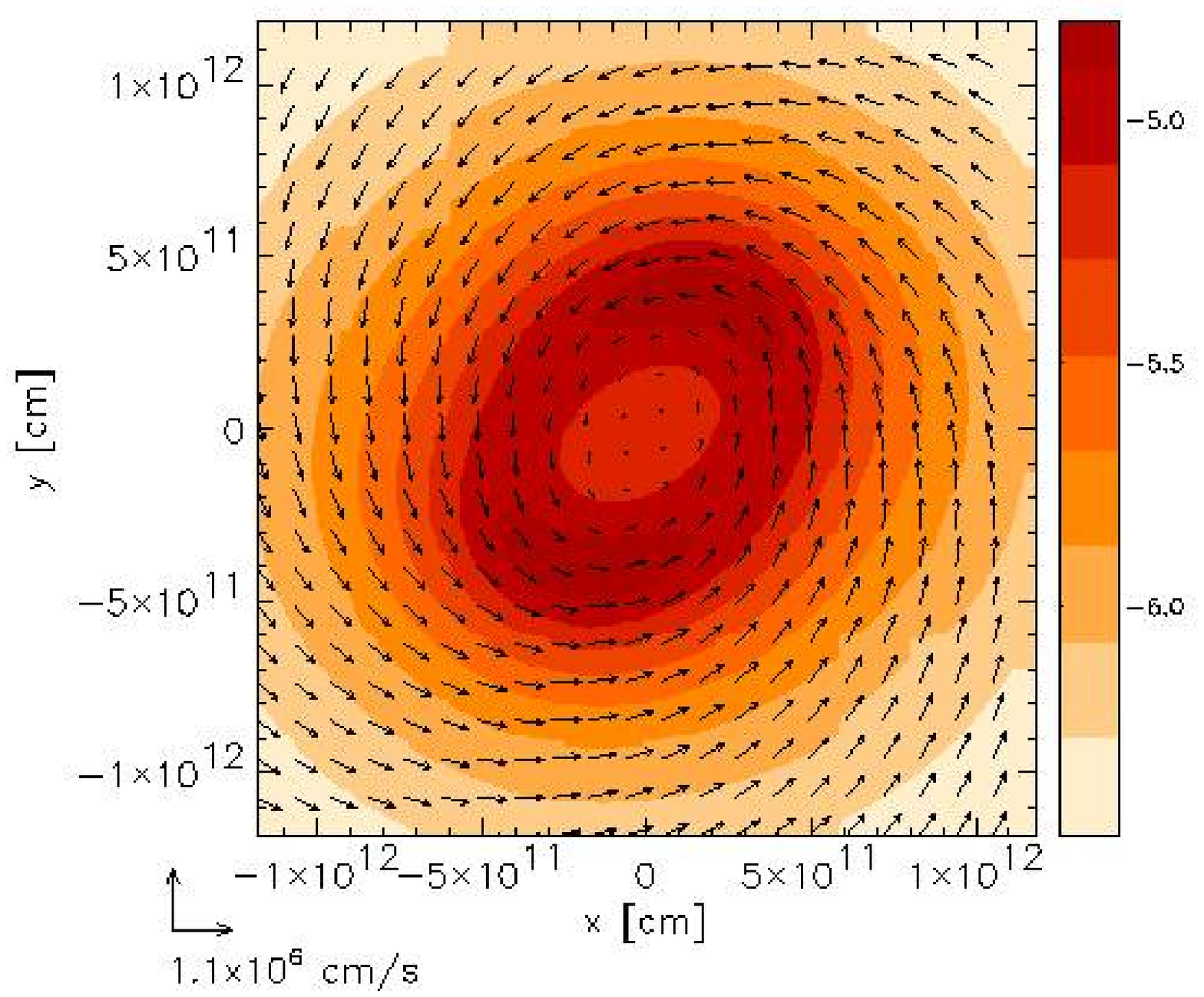} {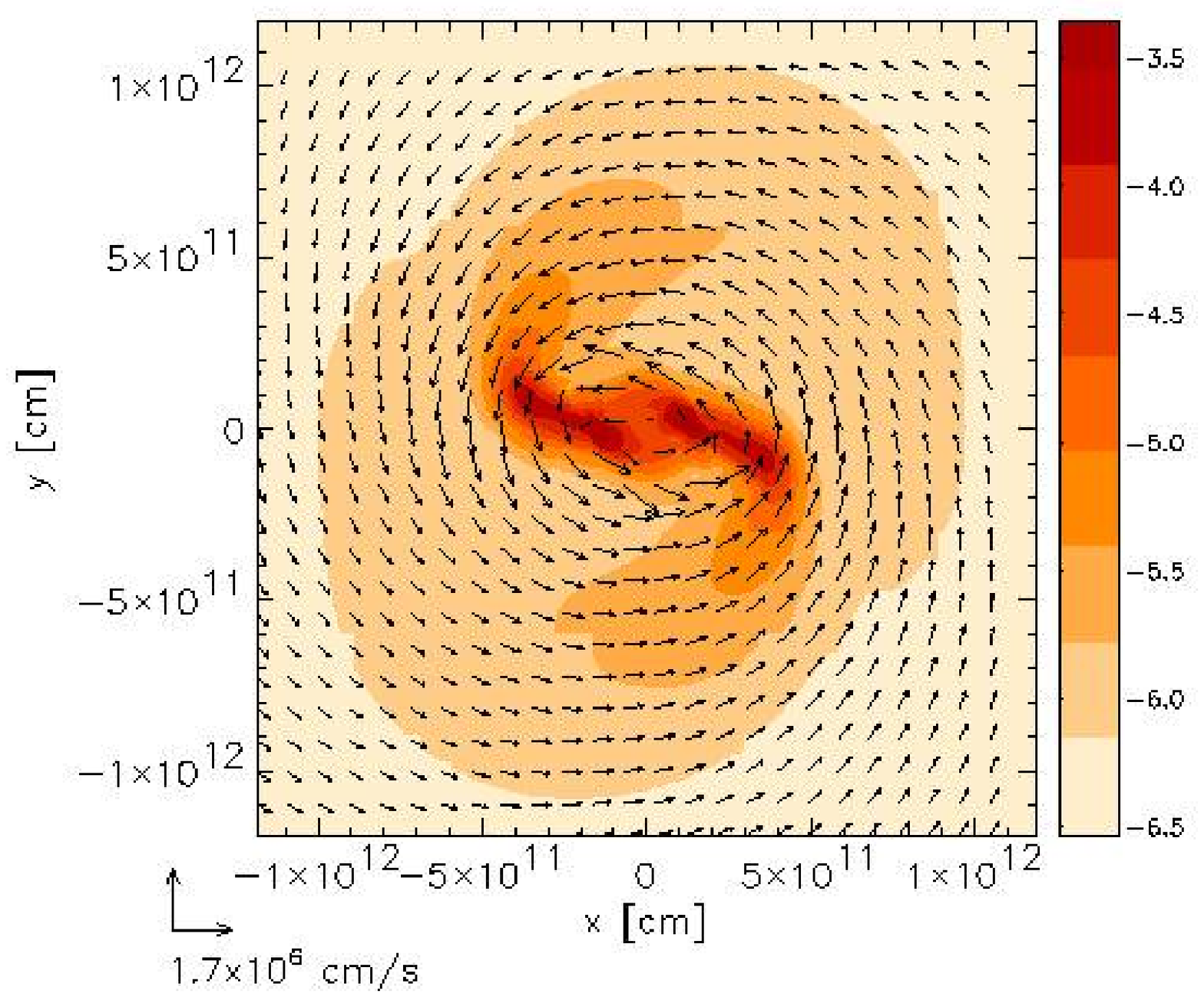}
\caption{Protodisk and Protostar(s). This figure shows the evolution of
   innermost part of the protodisk and the protostars $70,000 \, \ys$
   after the runaway collapse. The upper panels show the wind from the
   protostellar region in the plane perpendicular to the disk plane
   ($xz$), and the lower panels show the protodisk in the disk
   mid-plane ($xy$). An earlier formed ring structure (left panels)
   breaks up into two fragments after a few weeks (right panels). The
   color scale shows the gas density distribution (logarithmic scale
   in $\g \, \cm^{-3}$) and the arrows indicate the velocity
   field.}
\label{fig:protostar}
\eef

We examine accretion disk structure, infall, and jet outflow in the
vicinity of the protostar - on a physical scale of $0.07 \, \au$ and
smaller - in Figure~\ref{fig:protostar}. The snapshots in the left
panels are taken before the fragmentation of the disk into a ring has
taken place, while the right panels show the situation a few weeks
later, at the end of our simulation, after the disk's ring structure
has broken up into two pieces. This binary is only separated by $d
\sim 3 \, \Rsol$ and contains gas of $M_\star \sim 3\times 10^{-3} \,
\Msol$. These images are taken at an unprecedented level of grid
refinement as we able to study the dynamics on a physical scale of
$\sim 5\times 10^{10}\, \cm$. 

In the left panels, vigorous infall is still occurring onto the
central protostellar region as the material in the envelope rains down
on the central protostar. In the top right panel, a jet-like outflow
has been initiated.  This outflow achieves super Alfv\'enic speeds and
collimate towards the outflow axis. We show in the bottom right panel
that the smooth disk at this point undergoes gravitational
fragmentation into a binary protostellar system. The location of the
protostellar cores mark a transition of the jet pattern: gas inside
the binary system is falling back onto the protostars whereas material
outside the cores is leaving the disk. This bifurcation point
separates the disk dynamics from the forming binary.

From the upper right panel of Fig.~\ref{fig:protostar} one can see
that the disk wind arises from the circumbinary part of the disk. The
decrease in disk rotation speed $v_{\phi}$ inside $5\times 10^{11} \,
\cm$ may be a reason why the disk wind in this central region is
apparently more difficult to launch.

The disk, in the early stages that we are investigating here, is still
more massive than the protostars that are forming within it.  It is
therefore prone to gravitational instabilities that can lead to
gravitational fragmentation. The disk has fragmented into a ring of
material whose radius is of the order of $5 \times 10^{11}\,
\cm$. After one and a half rotations of the disk, the ring further
fragments into two pieces, each of which can be regarded as a
protostellar core in its own right. These fragments are strongly bound
by tidal interactions and there is the distinct possibility that they
will merge to form a single protostar since they are currently
separated by only a few solar radii.

Compared to our earlier -- pure hydrodynamical -- simulations~(\BPH),
we find that the physical scale of the ring fragment seen in this
hydro{\em magnetic} simulation is much smaller than the ring found in
the purely hydrodynamic calculations. This suggests that magnetic
fields support protostellar disks against fragmentation in the low
density regime at larger radial distances. The additional magnetic
pressure and the slow-down by magnetic braking reduce the possibility
of the protodisk to fragment. Recent simulations by~\citet{Hosking04}
(SPH) and by~\cite{Ziegler05} (AMR, grid code NIRVANA) also
point out that the influence of magnetic fields prevents
fragmentation. At higher densities (in our case $\sim 10^{19} \,
\cm^{-3}$) magnetic fields cannot prevent fragmentation and a ring
structure forms. This inner region ($\sim 5\times 10^{11} \, \cm$) is
rotating much faster than the outer envelope which makes it more
likely to form a ring.

Finally we show the link between toroidal and poloidal field structure
in the disk-binary system in Figure~\ref{fig:magfields_binary}. The
{\em toroidal} field on scales much larger than the binary system is
quadrupolar, because this component must vanish on both the mid-plane
of the disk as well as on the rotation axis. On scales characterizing
the binary, a much more complex structure is evident. First, the {\em
poloidal} field traces the density structure of the binary-disk system
(see also Fig.~\ref{fig:protostar}) due to flux-frozen compression of
the magnetic field. Second, the {\em toroidal} field (shown in gray
scale) peaks just outside of the binary system at $\sim 5\times
10^{11} \, \cm$ ($xz$-plane). This maximum in $B_{\phi}$ pushes gas
and therefore poloidal field lines toward the central rotation axis.

\bef
\plotfour{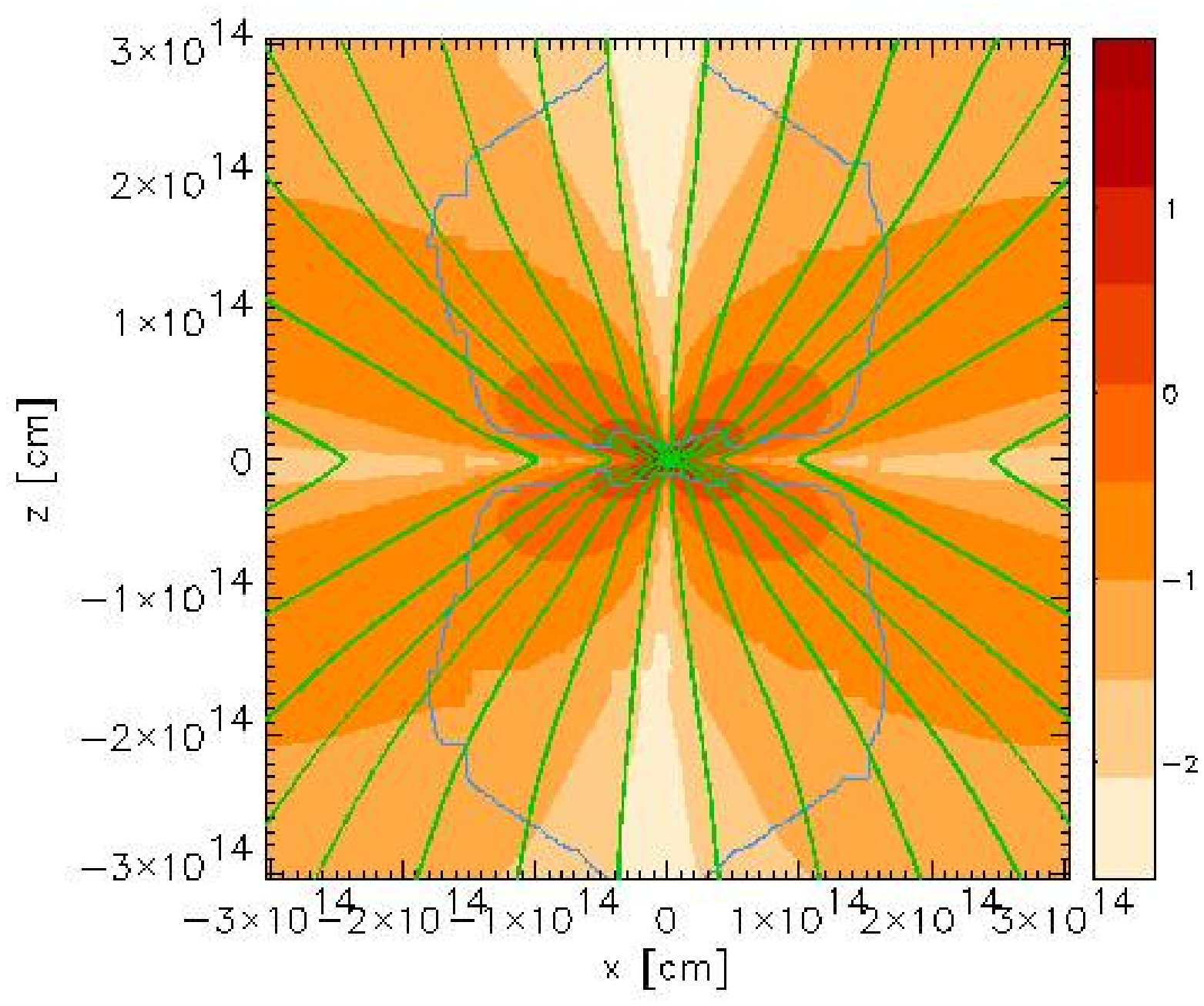}
         {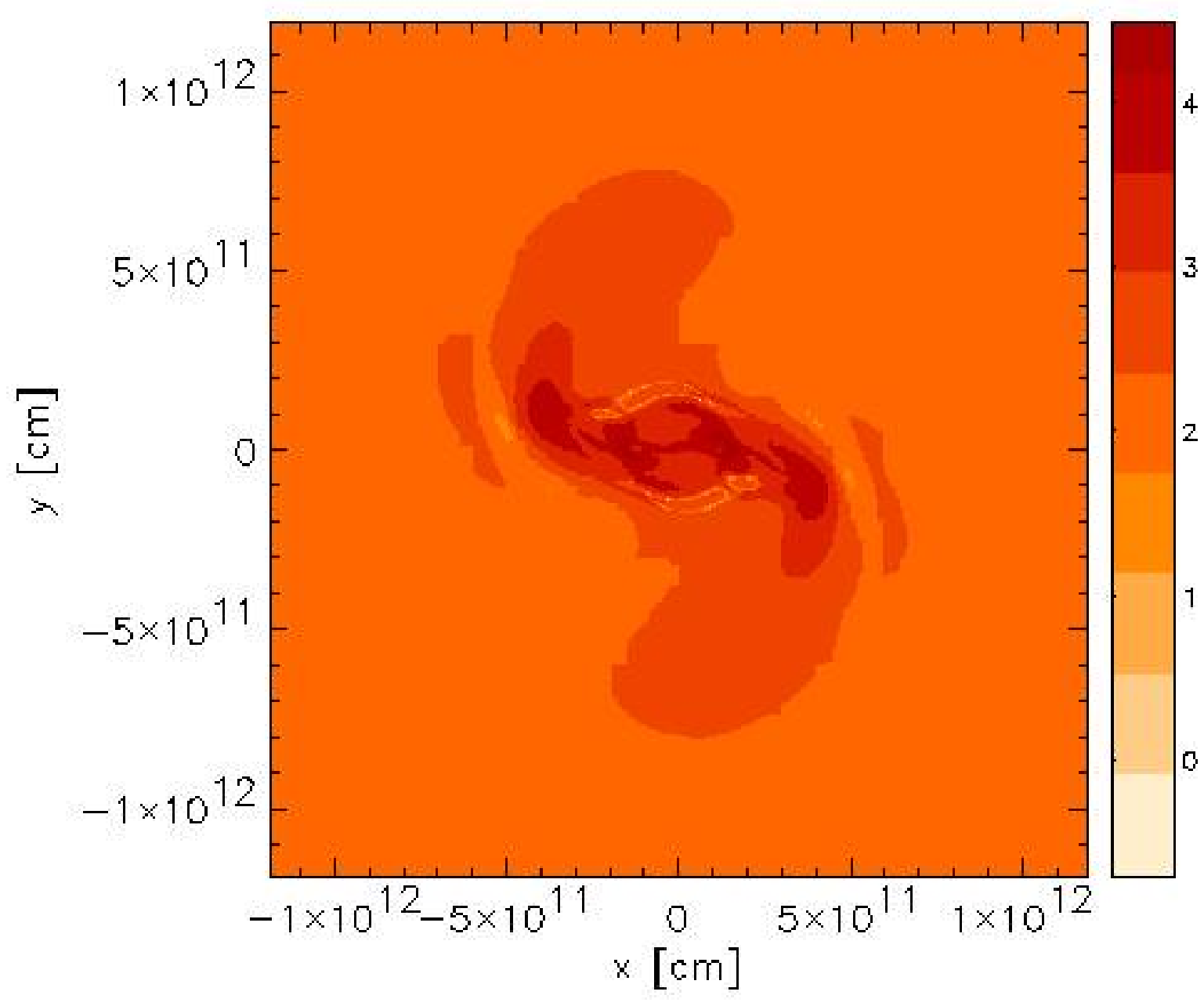}
         {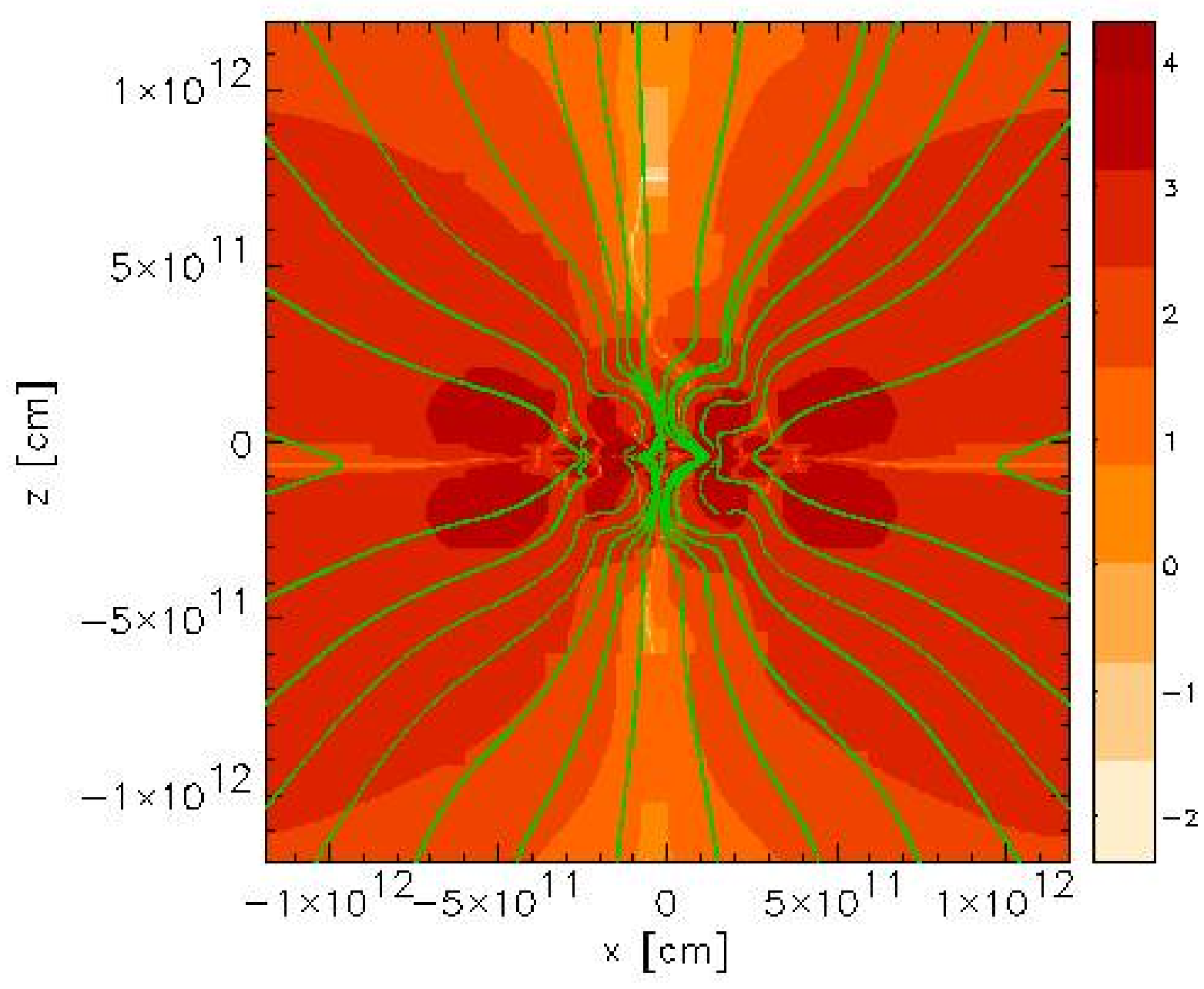}
         {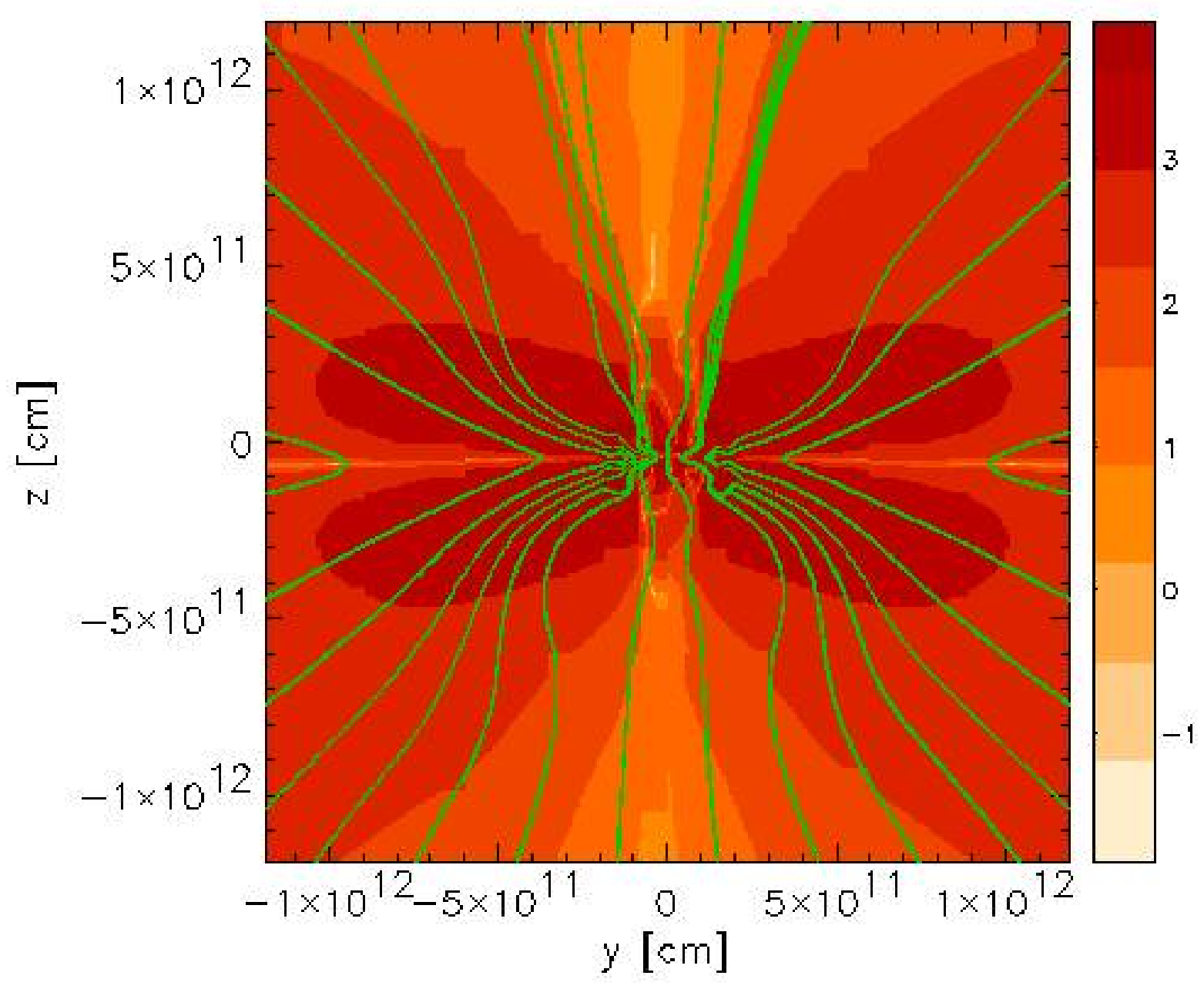}
\caption{Shows the magnetic field structure at the end of our
  simulation. The gray scale give the field strength in Gauss
  (logarithmic scale) and the 2D field lines are shown in green. (a)
  toroidal magnetic field at $\sim 20 \, \au$ scale, (b) poloidal
  field in the disk mid-plane, (c) toroidal field in the $xz$ plane,
  and (d) toroidal field in the $yz$ plane. The panels (b), (c), and
  (d) are snapshots with the same length scale than the protostellar
  disk of Fig.~\ref{fig:protostar}. Note that the slice (c) runs
  through the two proto-binary cores whereas slice (d) does not
  intersect either of these cores.}
\label{fig:magfields_binary}
\eef

\section{Properties of protostellar disks and magnetized protostars}
\label{sec:properties}

\bef
\plotfour{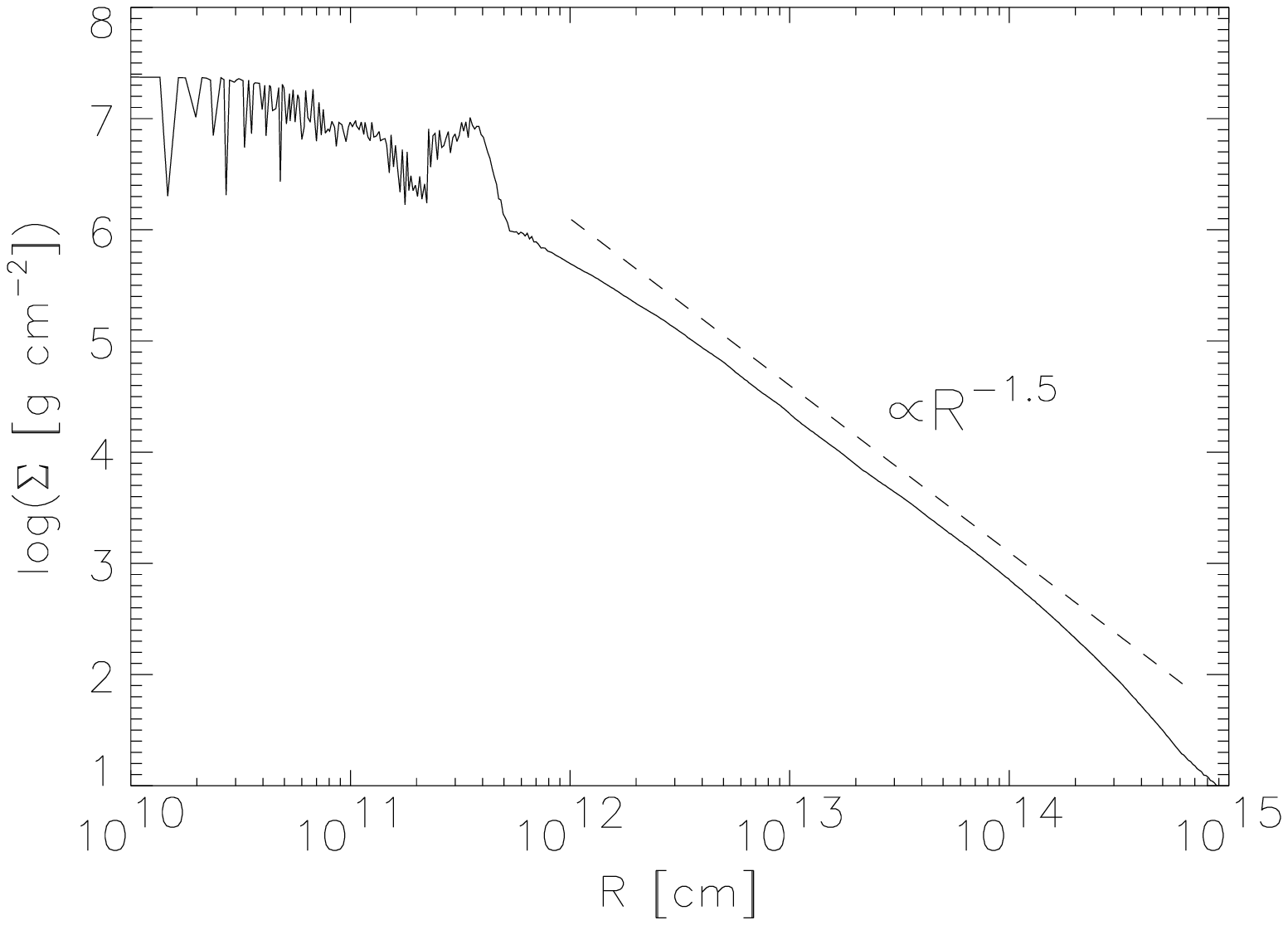}
         {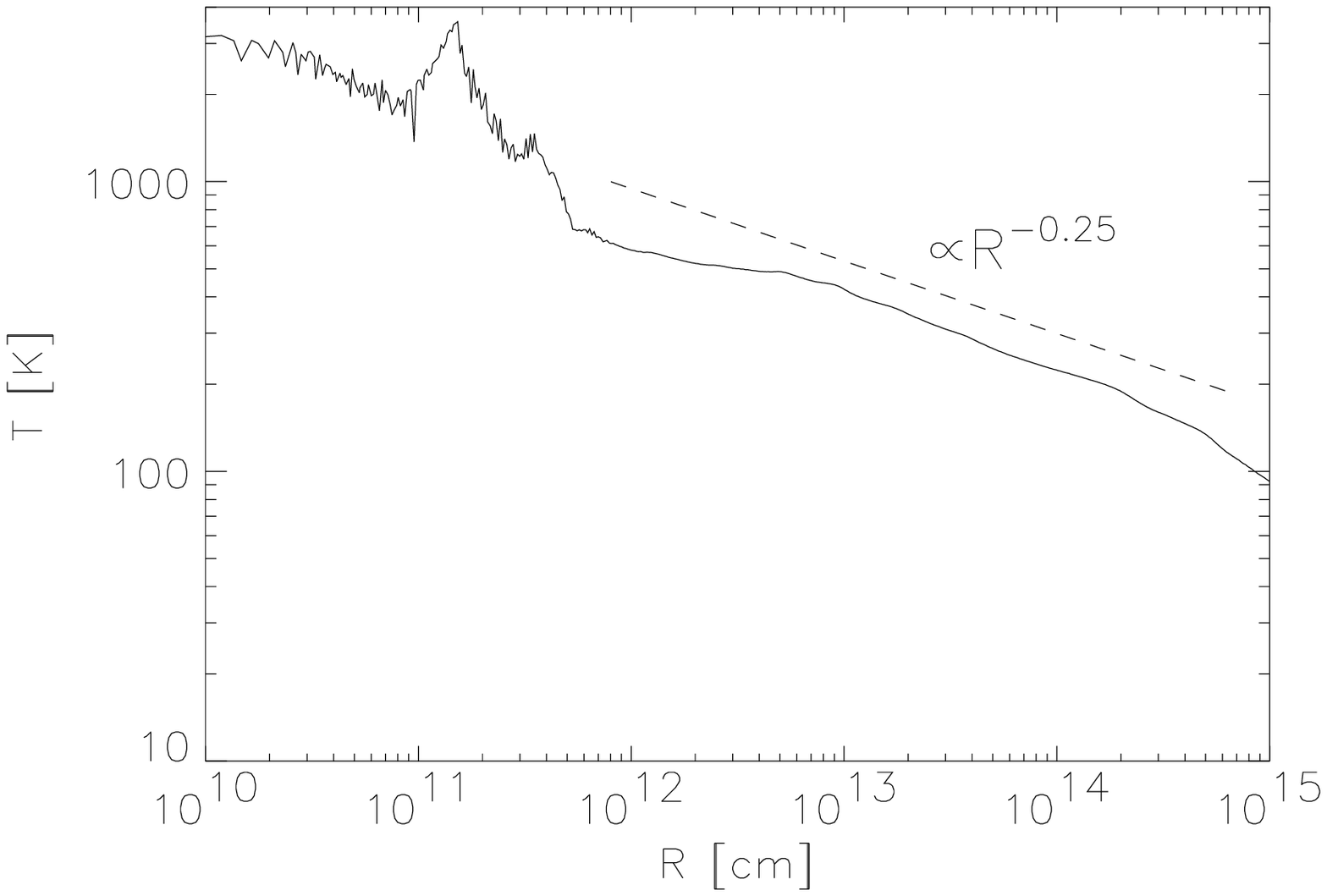}
         {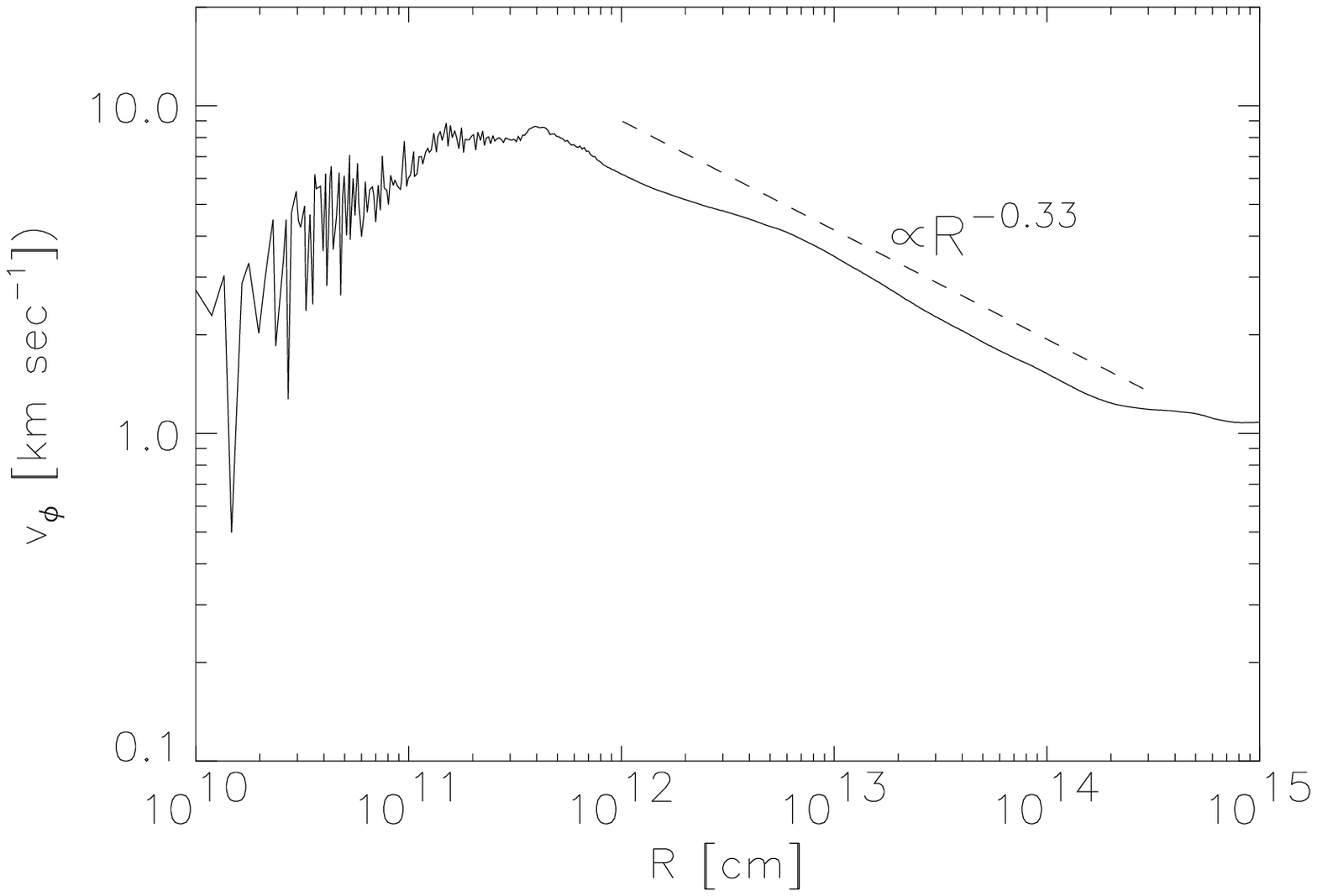}
         {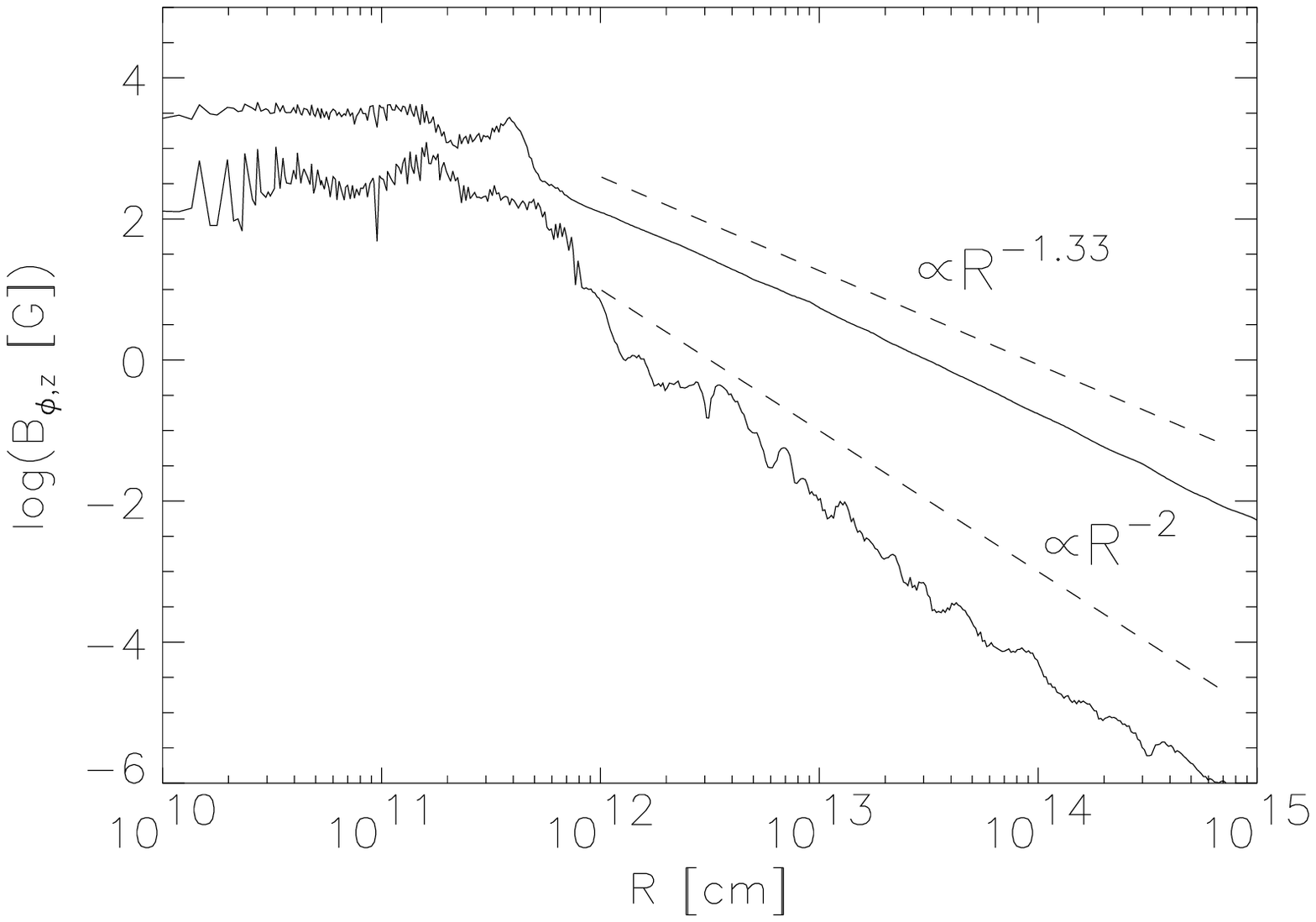}
\caption{Disk radial profiles. The panels show the radial
  (cylindrical) profiles of the column density, temperature, rotation
  velocity, and poloidal and toroidal magnetic field strength (from
  upper left to lower right) at the end of our simulation, i.e. $t =
  7\times 10^4 \, \ys$. The dashed lines show power laws indicated by
  the labels. The column density follows the Hayashi \citep{Hayashi81}
  law. We find that the toroidal component of the magnetic field,
  $B_{\phi}$, is smaller than the poloidal component, $B_z$,
  throughout the disk and follows a different power law. Note that the
  magnetic field trapped in the protostellar cores reaches fields
  strength of several $\kG$.}
\label{fig:profiles}
\eef

The radial structure of the disks that form in our gravitational
collapse simulations is of great interest for the observations of the
early stages of star and planet formation in real systems. Our results
pertain to the early (class 0) phases, before the central protostar
has become sufficiently massive to ignite the burning of its own
hydrogen fuel.  At this time the disk is heated primarily by the
energy that is liberated by shocks and adiabatic
compression. Figure~\ref{fig:profiles} consists of 4 panels in which
the radial profiles of various physical quantities of interest are
plotted as a function of the disk radius, $R$~\footnote{All quantities
are azimuthally averaged and weighted with the disk density, $\rho$,
e.g. $f(R) = (2\pi \, \Sigma(R))^{-1} \, \int \dd z \, \dd\phi \,
\rho(\vx) \, f(\vx)$, where $\Sigma(R) = (2\pi)^{-1} \, \int \dd\phi
\, \dd z \, \rho(\vx)$ is the column density as a function of the
cylindrical radius $R$.}.  These plots show that the disk's physical
properties can be very well described by simple power-law relations
over nearly three decades of physical scale. Interior to the smallest
scale is the region that is dominated by two protostellar cores.

The column density profile follows the famous "Hayashi" profile that
pertains for the structure of the disk in which our own solar system
may have been formed. It is interesting that the threading magnetic
field clearly plays a role in determining this structure since our
purely hydrodynamic collapse calculations resulted in a column density
profile that is considerably steeper. Models for the formation of
planetary systems often give values of physical quantities at
$1\,\au$, the position of the Earth. The column density at this early
time is $\Sigma (1\au) \simeq 10^4 \, \g \, \cm^{-2}$ which is
considerably denser than the Hayashi "minimum solar mass nebula" but
is still in agreement with values seen in young protostellar
disks \citep{Wilner00}. 

The disk, at $1\,\au$, has a temperature of $T(1\au) \simeq 400\,\K$,
which is somewhat higher than is found in most models.  We note that
this may be property of our assumed coolants that consist solely of
the excitations of molecules in the disks.  Real protostellar disks
will also be able to cool somewhat more efficiently by emission from
warm dust grains. The peak temperature achieved interior to a few
$10^{11}\, \cm$ reaches $3,000^{\circ} \, \K$, which is sufficient to
melt the dust grains in the gaseous disk~\citep{Hillenbrand92}. In our
case, this means that inside $1.3\times 10^{-3} \, \au$ interior to
the disk will be free of dust. Comparison of disk models with
observational data show that a typical disk will have an inner rim
that extends to $\sim 0.2 \, \au$ due to dust sublimation by stellar
and accretion heating~\citep{Muzerolle03}.

The rotation curve of a disk, whose gravitational field is dominated
by the central star, is given by Kepler's law, $v_{\phi,\sm{Kep}}
\propto R^{-0.5}$. In contrast, a fully self-gravitating disk has a
much shallower behavior with an angular velocity which is constant
with radius.  Our disk has an intermediate scaling, $v_{\phi}(r)
\propto R^{-0.33}$, which indicates that the disk's dynamics is
somewhat governed by the central forming protostar, but whose own
self-gravity is still significant. The rotation speed of material at
$1\,\au$ is $v_{\phi}(1\au) \simeq 3\,\km\,\sec^{-1}$.  This is
smaller by a factor of 10 than that of the Earth around our Sun.
However, at this moment in the simulation, this rotation speed is
completely consistent with the small amount of gas, as well as
protostellar mass, that is interior to $1\,\au$ - the former being
only $2.8 \times 10^{31}\,\g$, about 1 \% of the mass of the Sun.

Finally, the distribution of magnetic fields across the disk is a
vital ingredient in understanding how outflows are launched and how
the bulk of a disk's angular momentum may be extracted.  The dominant
field component is the vertical one, which scales as
\be
B_{z} \propto R^{-4/3} \, .
\ee
This radial dependence is close to the scaling one obtains for a
self-similar disk model wherein $B_z \propto R^{-5/4}$
\citep{Blandford82}.  Here the self-similarity scaling breaks down
however because the toroidal field component falls off much more
steeply than does that of the vertical field; we find the scaling
close to $B_{\phi} \propto R^{-2}$.
The strength of the dominant field component at $1\,\au$ is
quite significant; $B_z(1\au) \simeq 3.2$ Gauss.  It is remarkable
that meteorites found at about this physical scale in the solar system
are found to have been magnetized by a field strength that is of this
order - 3 Gauss \citep{Levy78}.

The vertical magnetic field in these data levels off in the innermost
region as is seen in this panel.  The peak field observed is of the
order of $3 \, \kG$.  This result is within the limits of measured
mean surface magnetic field strengths of stars, which are observed to
have values $\simeq 2\,\kG$ \citep{Johns-Krull99}. The interesting
conclusion is that the magnetic fields in protostars may be fossils of
this early star formation epoch in which the magnetic field of the
parental magnetized core was compressed into the innermost regions of
the accretion disks. Recently it was shown by~\citet{Braithwaite04}
that fossil fields with non-trivial configuration can survive over a
star's lifetime (see also our discussion in Sec.~\ref{sec:model}).

Recent measurements by~\citet{Donati05} of magnetic fields in the
accretion disk around FU Ori support many of our results reported
here. Firstly, the reported magnetic fields strength is around $1 \,
\kG$ at a distance of $0.05 \, \AU$ from the protostar. Secondly, the
inferred field configuration has the structure of a wound-up and
compressed magnetic field which azimuthal component points in the
opposite direction of the disk rotation and where the poloidal
component dominates the field. Thirdly, the magnetic plasma rotates
sub-Keplerian which indicates that strong magnetic braking must be
taking place.

\bef
\plotone{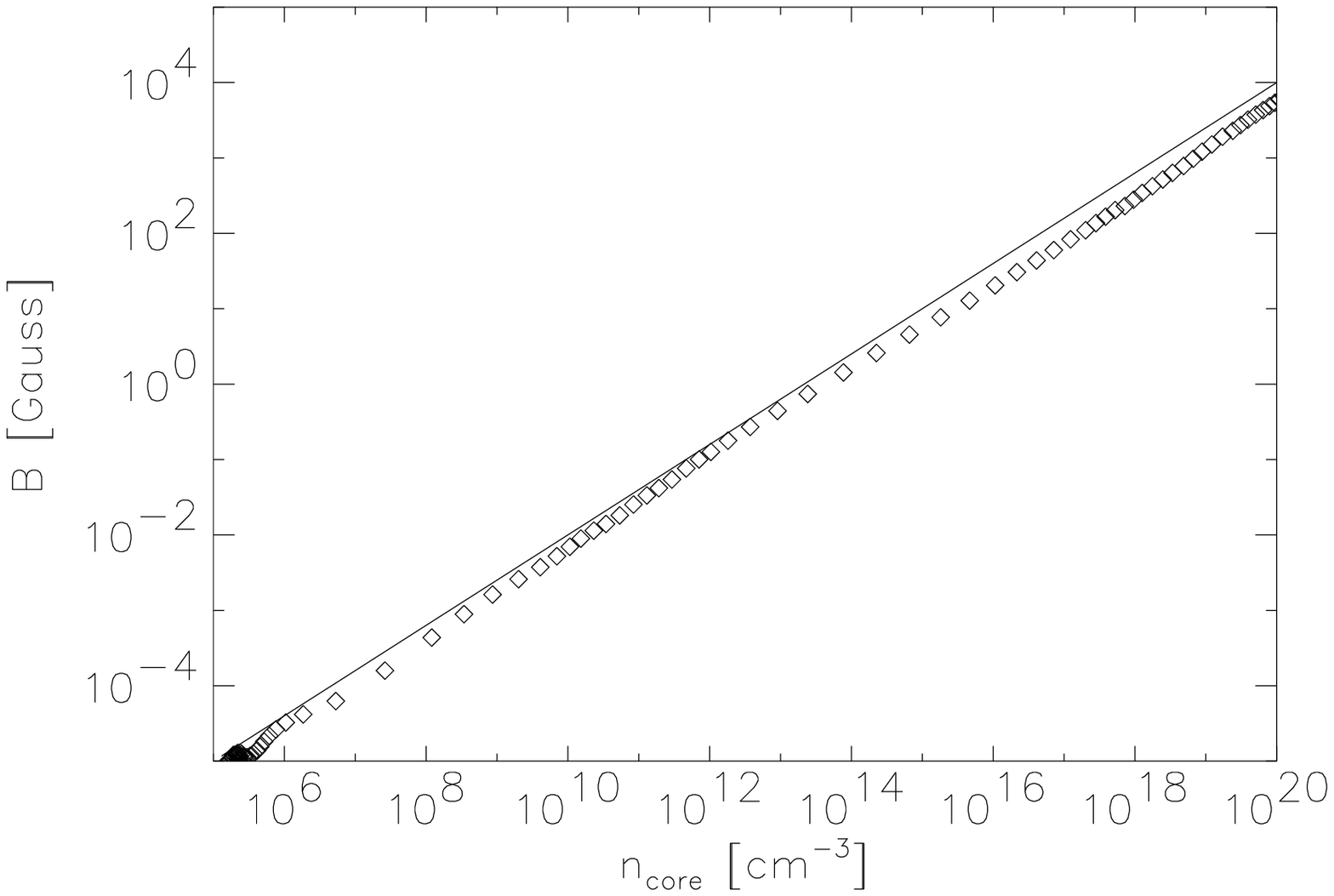}
\caption{The total magnetic field strength, $B$, as a function of the
  core density $n_{\sm{core}}$ from our simulation (diamonds). The
  magnetic field increases with the increase of the core density
  following a power law. We find a power law index of $\gamma \sim
  0.6$ ($B \propto \rho^{\gamma}$, solid line). This is slightly
  steeper than what is found in similar simulations without
  rotation \citep{Desch01} and core formation simulations \citep{Li04}
  which found $\gamma \approx 1/2$. Note that the core density is a
  increasing function of time and each data point is calculated at
  different times.}
\label{fig:mag_by_dens}
\eef

Figure~\ref{fig:mag_by_dens} summarizes the evolution of the magnetic
fields strength as a function of the core density $n_{\sm{core}}$. We
find the scaling relation $B \propto n^{0.6}$. This power law is
slightly steeper than is found in other simulations
\citep[e.g.][]{Desch01, Li04} who found $B \propto n^{1/2}$ in
collapse and core formation calculations. In general, the scaling law,
$B \propto n^\gamma$, depends on the field geometry. For instance,
small scale tangled magnetic fields scale as $B \propto n^{2/3}$. In
our case, we have poloidal {\em and} toroidal field components which
lead to a slightly steeper compression of the field strength with
density~\footnote{We calculate the mean magnetic field strength, $B =
\sqrt{\left<B^2\right>}$, as follows: $\left<B^2\right> = V^{-1} \,
\int \dd V \, \left(B_x^2 + B_y^2 + B_z^2\right)$, where the volume,
$V$, is taken in the region where $n_{\sm{max}}/2 \le n \le
n_{\sm{max}}$, $n_{\sm{max}}(t)$ being the maximal density at the
time, $t$.}

The origin of the magnetic fields in accretion disks remains one of
the most important, unresolved issues of star formation. Fields could
be advected inwards during the collapse of the magnetized core on
larger scales -- as we see in our simulations. It is also
theoretically possible for magnetic fields to be generated within
disks by dynamo action~\citep[e.g.,][]{Pudritz81a, Pudritz81b,
Stepinski88, Brandenburg95, Rekowski03}. The simulations
of~\citet{Rekowski03} explicitly include a dynamo generation term
(``$\alpha$'') in the induction equation. Their results show that
dynamo action in a turbulent disk can generate fields as well as disk
winds without the need for larger scale external fields. In our
simulations however, the time span is not sufficient to observe dynamo
generated disk fields as we can follow only a few disk rotations in
this collapse situation.

\bef
\plotone{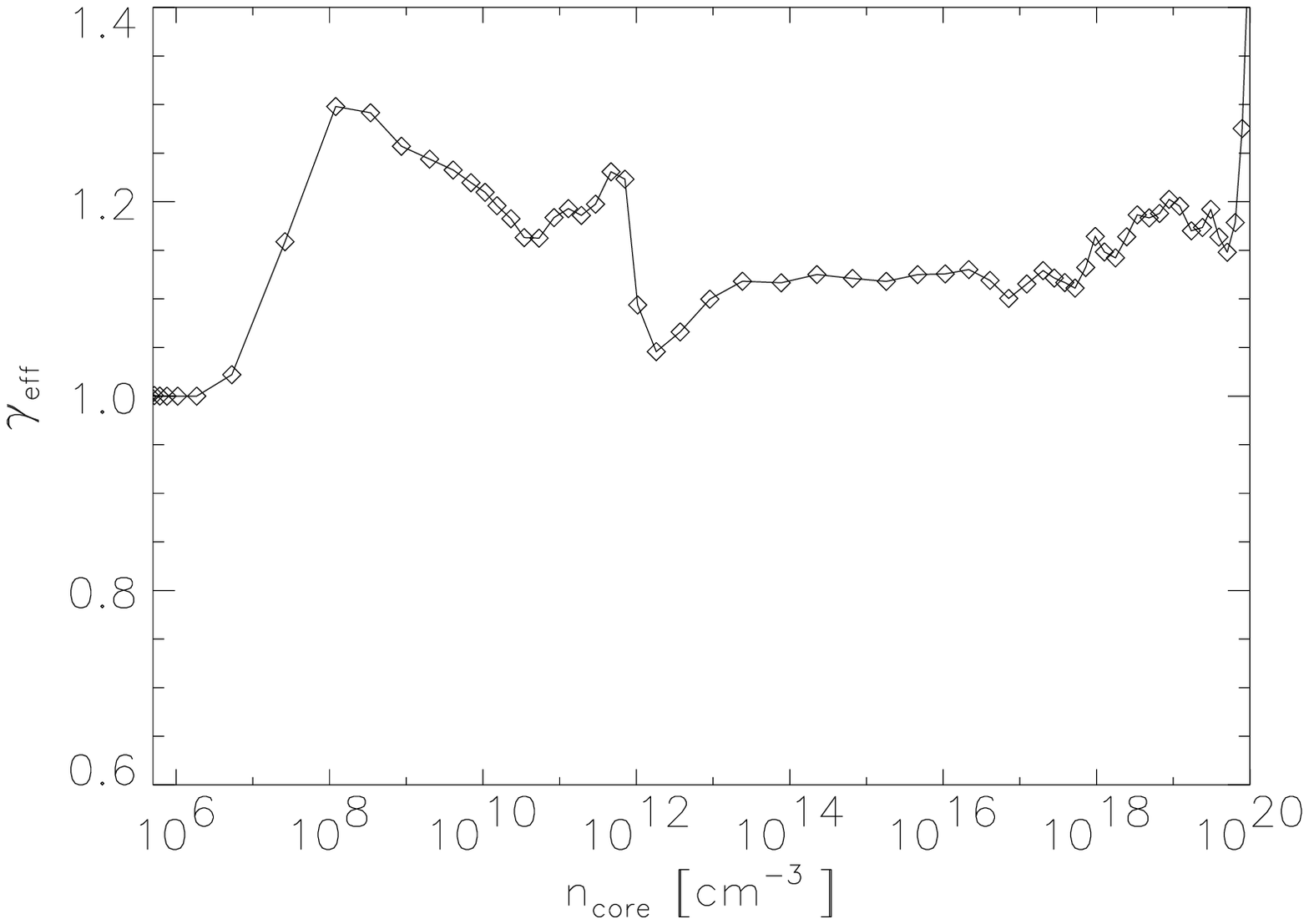}
\caption{The effective equation of state, $\gamma_{\sm{eff}} =
  \dd\ln{p}/\dd\ln{\rho}$ as a function of the core density
  $n_{\sm{core}}$. Cooling by molecular line emission is effective
  until $n \sim 10^{7.5} \, \cm^{-3}$ above which the effective EOS
  reflects the molecular cooling ability in the different
  density/temperature regimes as well as the appearance of shocks.}
\label{fig:EOS}
\eef

In Fig.~\ref{fig:EOS} we show the evolution of the effective EOS,
$\gamma_{\sm{eff}} = \dd\ln{p}/\dd\ln{\rho}$, in the core region. We
performed the same analysis for our purely hydrodynamical simulations
(\BPH) where $\gamma_{\sm{eff}}$ varies much more significantly with
time. For these simulations we did not have as many output files,
i.e. the analysis relied on a poorer time resolution.  Note in
particular the significant change in $\gamma_{\sm{eff}}$ in the range
of densities $10^8 - 10^{13} \, \cm^{-3}$. The various changes of the
effective EOS during the collapse phase reflect the necessity to
include realistic physics into collapse simulations and cast some
doubt on the use of simple adiabatic ``switching'' to study gas at
high densities. The relevance of cooling for dynamical collapse
simulations is also pointed out by~\citet{Lesaffre05}.

\section{The disk-jet connection}
\label{sec:disk-jet}

\bef
\plotfour{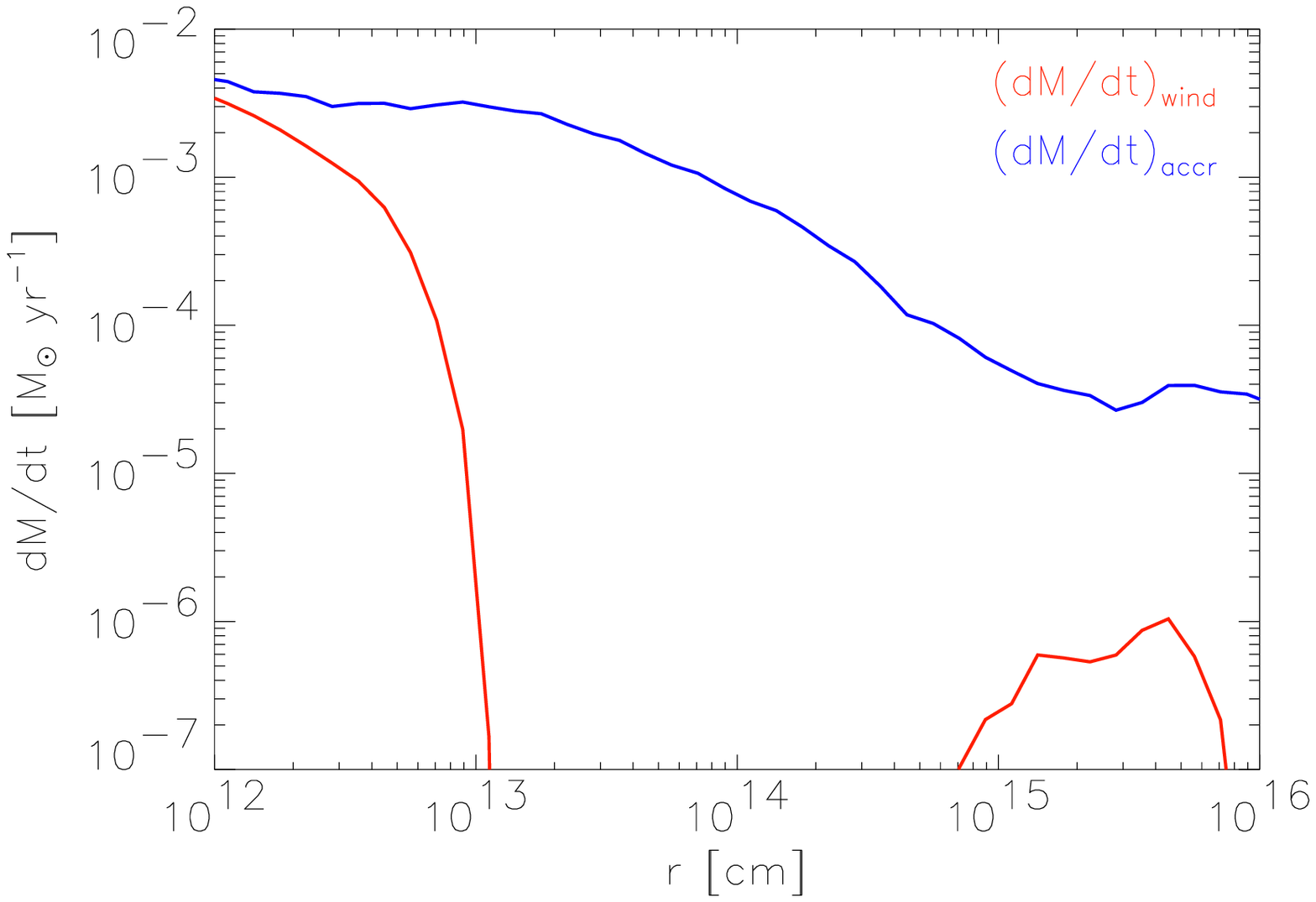}
         {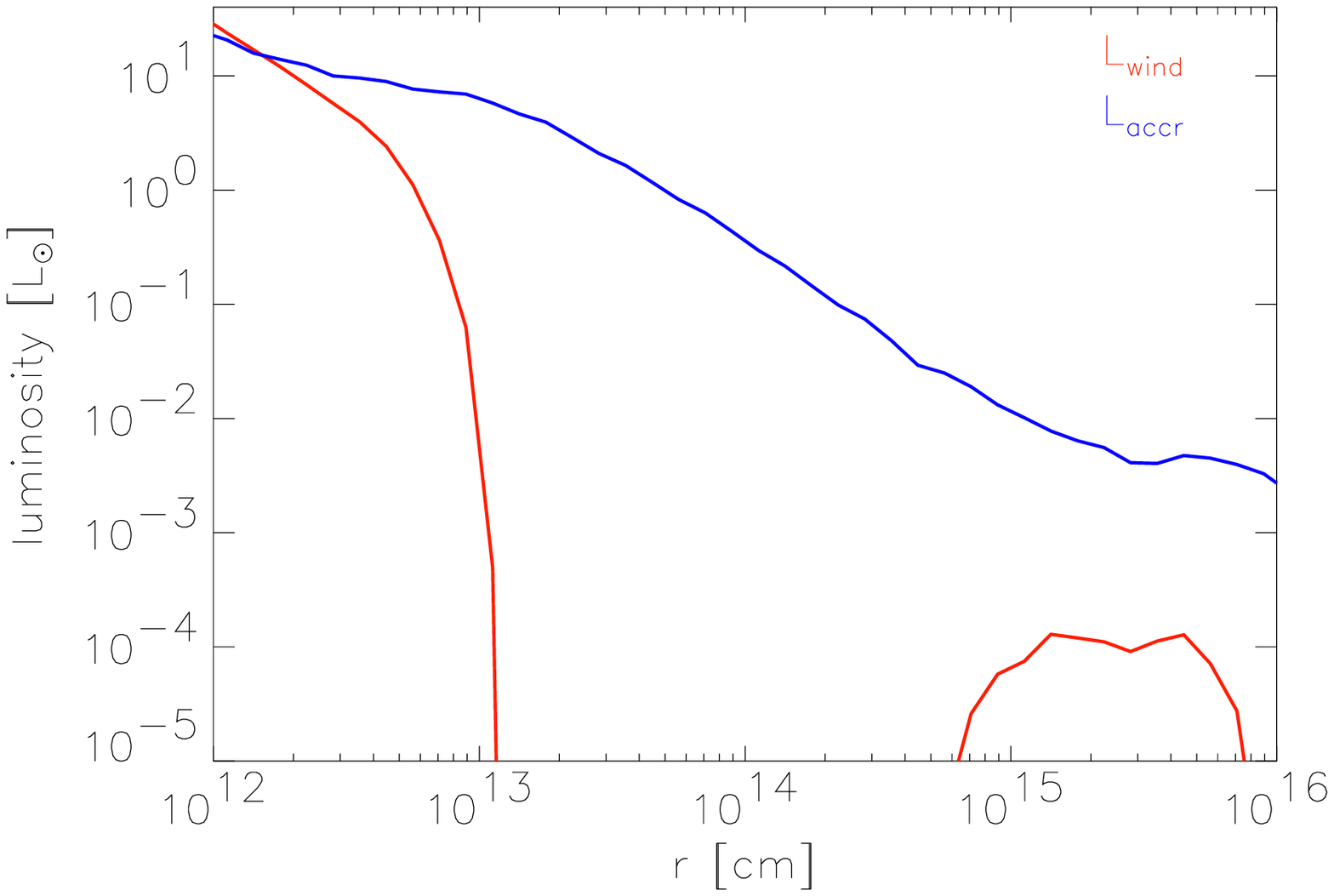}
         {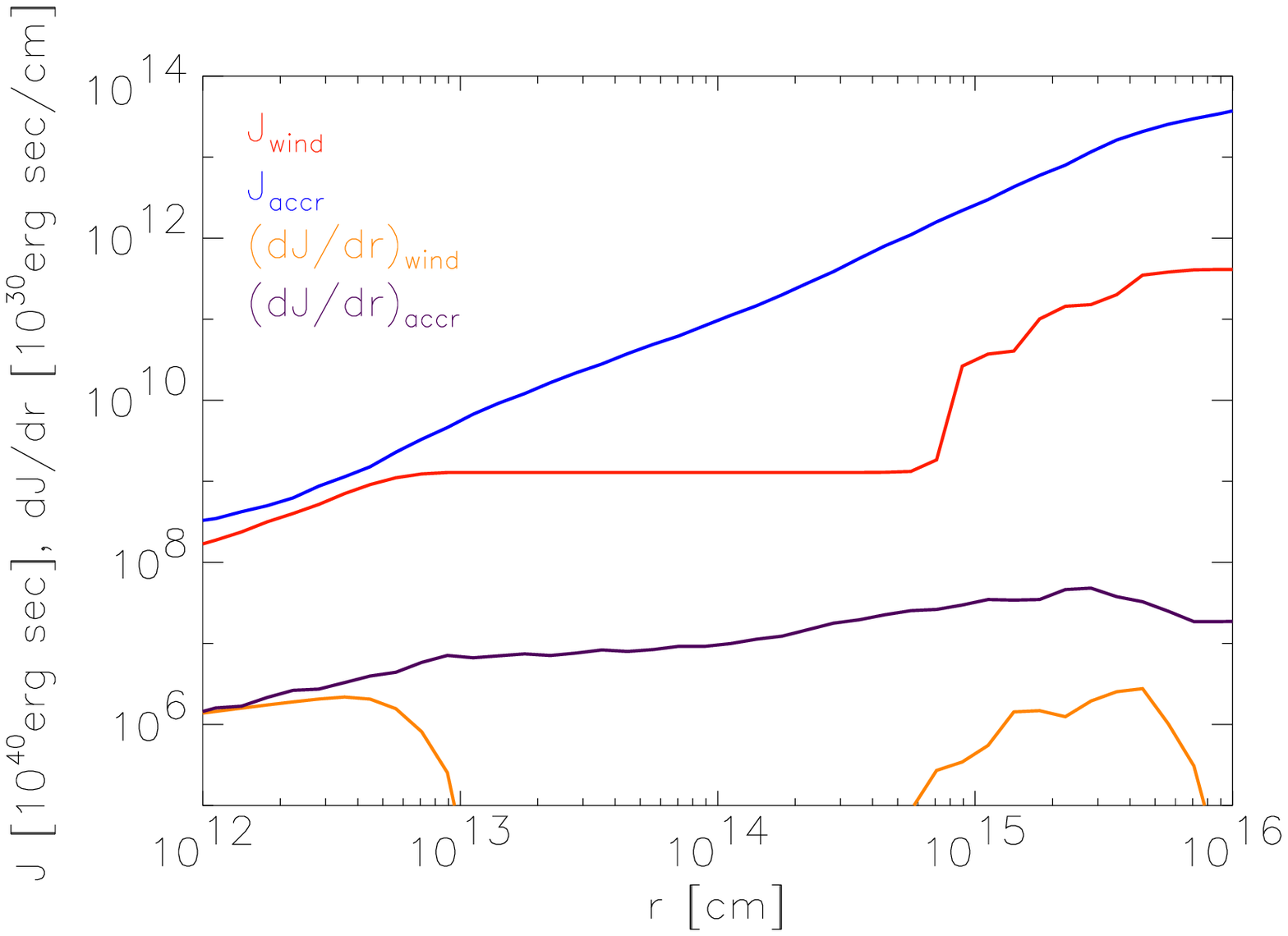}
         {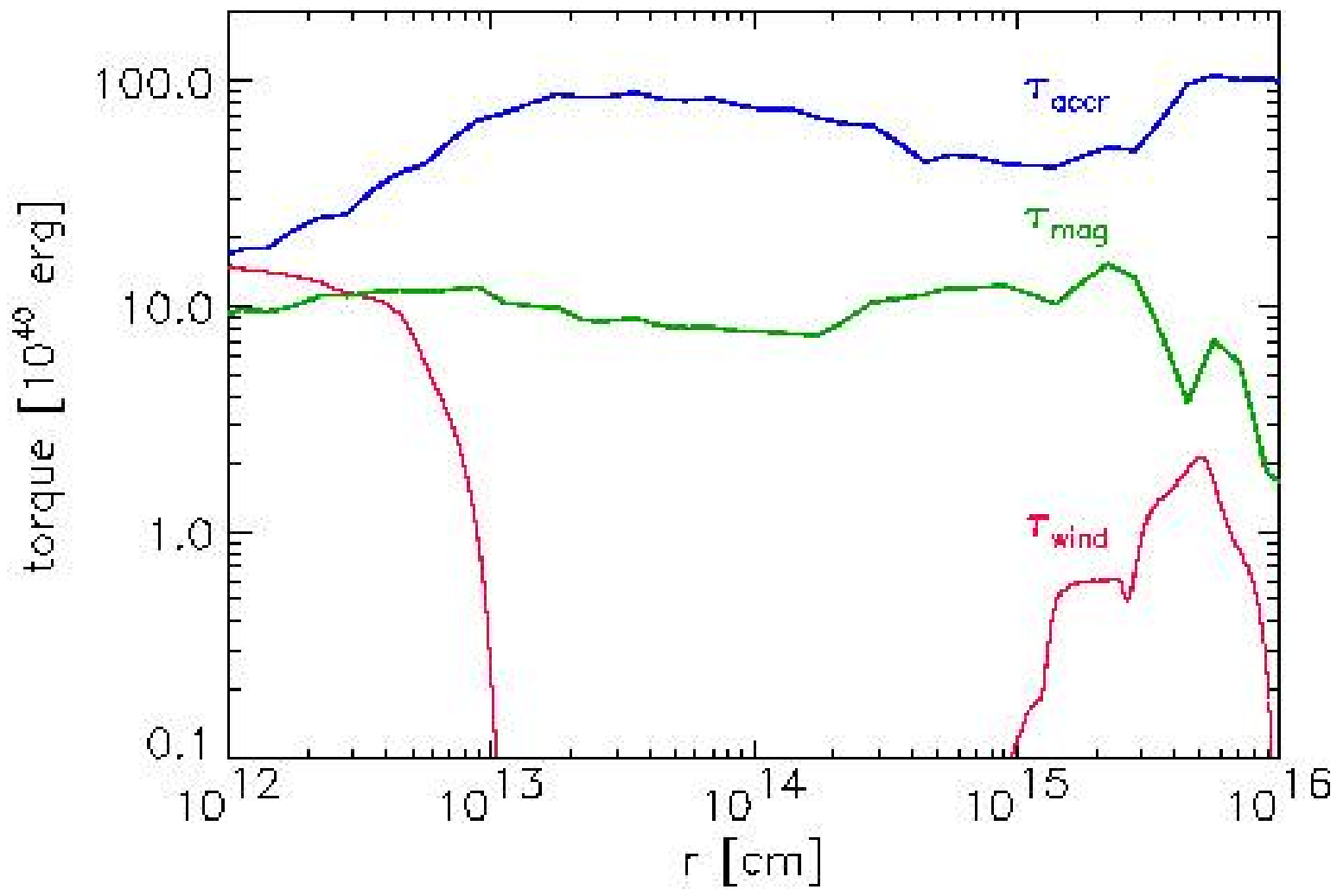}
\caption{Disk-Wind Connection. These panels show the properties of the
  disk-wind connection at the end of our simulation, i.e. $t = 7\times
  10^4 \, \ys$. From top left to bottom right: mass outflow and mass
  accretion through the disk plane; kinetic luminosities from the wind
  material and gas ``raining'' onto the disk; angular momenta, $J$,
  carried by the wind and by the disk; total torques (change of
  angular momenta) of the magnetic field, the wind and accretion
  disk. The double shock structure at $r \sim 1\,\au$ and $r \sim
  600\,\au$ is clearly visible in the wind properties: the jet and the
  large scale outflow are still confined beneath these shock
  fronts. Both, the jet and the large scale outflow, are launched at
  different disk heights by different driving mechanisms (see
  text).}
\label{fig:diskjet}
\eef

Outflows and jets from young stellar objects are powered by the
accretion flows in the disk where magnetic fields redirects a small
fraction of the flow leading to a mass loss from the
protodisk. Therefore, accretion properties and wind properties should
be strongly correlated. Observations of various different systems
indicate for instance that the mass loss by the wind
$\dot{M}_{\sm{wind}}$ and the mass accretion through the disk
$\dot{M}_{\sm{accr}}$ are related by
$\dot{M}_{\sm{wind}}/\dot{M}_{\sm{accr}} \sim
0.1$ \citep[e.g. review][]{Konigl00}. In Fig.~\ref{fig:diskjet}, we
summarize the results of our simulation which compares the wind
properties with that from the accretion disk $70,000 \, \ys$ after the
onset of the gravitational collapse (quantities are shown as averaged
over a sphere with radius $r$). Clearly visible in these graphs are
the double shock structures that confine the jet and the large scale
outflow to distinct regions. The inner shock ($\sim 1\,\au$)
momentarily hinders the jet from expanding towards higher disk
altitudes and the outer shock ($\sim 600\,\au$) until now blocks the
outflow from running into the cloud envelope. Nevertheless, the jet is
already powerful enough to expel material at a rate of
$\dot{M}_{\sm{jet}} \simeq 10^{-3} \, \Msol / \yr$ at $r = 3
\times 10^{12} \, \cm$ which is a good fraction of the peak accretion
through the disk ($\dot{M}_{\sm{jet}}/\dot{M}_{\sm{accr}} \sim 1/3$).

The system has not yet reached a steady state configuration and is
still in a contracting phase. Therefore, the large scale outflow is
not yet fully developed and its contribution to the mass loss and
luminosity is still gaining importance. Nonetheless, we can classify
the last stage of our simulation as a ``Class 0'' protostar because the
luminosity of the large scale outflow might be in the observable range
(a few \% of the accretion luminosity). 

In the lower left panel of Fig.~\ref{fig:diskjet} we show the angular
momentum per unit length,
\be
\frac{\dd J^{\pm}}{\dd r} = r^2 \, \int \dd \Omega \,
 \rho \, R \, v_{\phi} \, ,
\ee 
and the total angular momentum
\be
J^{\pm} = \int \dd r \, \frac{\dd J^{\pm}}{\dd r} \, ,
\ee
carried by the infalling (accretion, $-$) and outgoing (wind, $+$)
material, where $v_{\phi}$ is the toroidal velocity and $R$ is the
cylindrical radius. At this early times the small scale jet already
carries a significant amount of the total angular momentum which is
extracted from the protodisk.

The lower right panel of Fig.~\ref{fig:diskjet} shows the torques
exerted by the accretion flow, wind, and magnetic field, where we
calculate the accretion torque ($-$) and the wind torques ($+$) as the
angular momentum flux through a sphere with radius $r$,
\be 
\tau^{\pm} = r^2 \, \int \dd \Omega \, v^{\pm}_r \, 
      \rho \, R \, v_{\phi} \, ,
\ee
where $v^{\pm}_r$ is the radial outflow/inflow velocity. The torque
exerted by the magnetic field on the fluid at radius $r$
is \citep[e.g.,][]{Konigl00} 
\be
\tau_{\sm{mag}} = \frac{1}{4\pi} \, r^2 \, \int \dd \Omega \, 
   \left(B_r + B_p\right) \, R \, B_{\phi} \, ,
\ee
where $B_r$, $B_p$, and $B_{\phi}$ are the radial, poloidal, and toroidal
components of the magnetic field.

The distribution of the torques indicate that the jet transports
almost as much angular momentum away from the inner part of the
protodisk as the disk gains from accreting gas. Together with the
angular momentum extracted by the magnetic torque from this inner
region, we conclude that the protostar(s) will spin much below its
break-up rate at the time the system relaxes to a steady state
configuration.

\section{Conclusion}
\label{sec:summary}

Our study on the early phases of gravitational collapse and star
formation in low mass, magnetized, molecular cores shows that outflows
and jets are an inevitable consequence of disk formation. Outflows
play a central role in transporting disk angular momentum, and in
determining the disk structure and basic properties of the forming
protostar(s). The predicted properties for the jets, namely that they
rotate and transport the bulk of the disk's angular momentum, have now
been confirmed by Hubble Space Telescope observations
\citep{Bacciotti02}.

Our study is a comprehensive and self-consistent approach -- including
cooling by molecular line emission -- of generating outflow phenomena
during the collapse of molecular cloud cores. However, additional
physical processes such as dust cooling and ambipolar diffusion need
to be added to our model (see also discussion in
Sec.~\ref{sec:model}).

Among our many findings, we list the most significant results below.
\begin{itemize}
\item We confirm in our simulations that angular momentum loss by the
  flux of torsional Alfv\'en waves can significantly reduce the spin
  of pre-collapse cores.

\item A large scale outflow (up to $600 \, \au$) occurs which is
  driven by the wrapped-up toroidal field pressure resulting in a
  magnetic tower flow. This creates a torus-like ``bubble'' that
  expands into the surroundings.

\item An inner jet (up to $0.7 \, \au$) is driven from the disk in the
  deepest part of the gravitational well and is powered by
  magneto-centrifugal force. This disk wind is launched from the disk
  region exterior to the binary.

\item The magnetic field suppresses, but does not entirely prevent ring
  fragmentation of our disk. The ring has a physical scale of $\sim 0.04
  \, \au$ and rapidly fragments into a proto-binary with a separation
  of only $\sim 3 \, \Rsol$.

\item The protostellar disk still dominates the mass of the forming
  protostars in these early ($t \simle 7\times 10^4 \, \ys$) stages
  with $M_{\sm{disk}} \approx 0.1 \, \Msol$ within $100 \, \au$
  whereas the binary mass is $M_\star \sim 3\times 10^{-3} \, \Msol$. 

\item The accretion rate through the disk varies strongly in both
  space and time. In the inner regions, we find
  $\dot{M}_{\sm{jet}}/\dot{M}_{\sm{accr}} \sim 1/3$ with a peak
  accretion rate in the disk reaching $\dot{M}_{\sm{accr}} \sim
  3\times 10^{-3} \, \Msol / \yr$.

\item During the collapse, the total magnetic fields strength scales with
  the core density as $B \propto n^{0.6}$.

\item Our results suggest that it is possible that stars acquire $\sim
  10^3 \, \G$ fields as fossils of this early collapse era, rather than
  through dynamo action on an initial weak field.
\end{itemize}

\acknowledgments

The authors thank Sean Matt for his valuable comments on this
paper. We like to thank Nick Gnedin for providing his
visualization tool, {\em IFRIT}, to the science community which
allowed us to produce the 3D pictures in this paper. We also thank an
anonymous referee for their constructive comments. The FLASH code
was developed in part by the DOE-supported Alliances Center for
Astrophysical Thermonuclear Flashes (ASCI) at the University of
Chicago. Our simulations were carried out on a 128 CPU AlphaServer SC,
which is the McMaster University node of the SHARCNET HPC Consortium.




\end{document}